\documentclass[a4paper,prb,twocolumn,floatfix,showpacs]{revtex4-1}
\usepackage{graphicx,amsmath,pstricks}

\begin{document}


\hspace{5.2in}
\mbox{Fermilab-Pub-04/xxx-E}

\title{Dangling bonds and magnetism of grain boundaries in graphene}

\author{M.A.~Akhukov$^1$}
\author{A.~Fasolino$^1$}
\author{Y.N.~Gornostyrev$^2$}
\author{M.I.~Katsnelson$^1$}

\affiliation{$^1$Radboud University Nijmegen/Institute for Molecules and Materials, Heyendaalseweg 135, NL-6525AJ Nijmegen, The Netherlands}
\affiliation{$^2$Institute of Quantum Materials Science, Ekaterinburg 620175, Russian Federation}

\date{\today}

\begin{abstract}
Grain boundaries with dangling bonds (DBGB) in graphene are studied by atomistic Monte Carlo and molecular dynamics simulations in combination with density functional (SIESTA) calculations.
The most stable configurations are selected and their structure is analyzed in terms of grain boundary dislocations. 
It is shown that the grain boundary dislocation with the core consisting of pentagon, octagon and heptagon (5-8-7 defect) is a typical structural element of DBGB with relatively low energies. 
Electron energy spectrum and magnetic properties of the obtained DBGB are studied by density functional calculations.
It is shown that the 5-8-7 defect is magnetic and that its magnetic moment survives after hydrogenation. 
The effects of hydrogenation and of out of plane deformations on the magnetic properties of DBGB are studied.
\end{abstract}

\pacs{61.48.Gh, 73.22.Pr, 61.72.Mm, 75.75.-c}


\maketitle

\section{Introduction}
\label{sec:introduction}

Most potential applications of graphene require to construct macroscopically large samples that are bound to be polycrystalline. Several routes are currently actively pursued to obtain large samples in an efficient way. Examples are evaporation of surface layers of SiC \cite{Berger1,Berger2}, solution of graphite without functionalization, in combination with sonication \cite{14,15,16} to obtain graphene paper (laminate) and chemical vapor deposition on metals \cite{LiScience,Bae}. Studies of graphene grown by these methods confirm the existence of grain boundaries (GB), as was observed in graphene on SiC \cite{11Y}, Ir(111) \cite{12Y,13Y}, polycrystalline Ni \cite{14Y} and Cu \cite{mceuen}. Although the presence of GB may be detrimental for electron mobility and mechanical strength, GB are potentially interesting by themselves, e.g. by metallicity along the grain as shown in Ref. \cite{Lahiri}. Several theoretical papers have considered the structure \cite{Yazyev,Grantab,Malola,Carlsson} and electronic \cite{Yazyev2} properties of tilt GB in graphene.

GB were subject of intensive experimental and theoretical study in the 70's of last century. At that time, the basic principles of formation of GB structures were understood and the special class of GB characterized by high symmetry was identified by the coincidence site lattice (CSL) approach \cite{CSL}. These GB have optimal matching of the grains and, being energetically the most favorable, are dominant in well annealed polycrystalline samples. Most GB studied experimentally in graphene can indeed be classified as low energy structures within the CSL theory \cite{Carlsson,Bae}. These GB consist of regularly arranged dipoles of disclinations with rotation angles $\pm 60^{\circ}$ associated with 5 and 7-fold carbon rings \cite{Yazyev}. The distance between disclination dipoles depends on the misalignment of the grains. The high strength characteristics of these GB in graphene \cite{Grantab} confirms the strong bonding in the core of the 5-7 disclination dipoles.

In bulk materials, however, also less favorable GB with extrinsic structural defects, extra volume excess and large elastic strain have been observed depending on the treatment of polycrystalline samples \cite{Valiev}. Also for graphene, one may expect this situation for samples obtained by coalescence of  independently growing nuclei as typical of chemical vapor deposition. The properties of more general GB have been considered in Ref. \cite{Mesaros} and in Ref. \cite{Malola} it was shown that, besides 5-7 pairs, there are 8-fold rings which dominate at tilt angle close to 15$^{\circ}$ as well as 4- and 9-fold rings with less probability. Beside having higher energy and excess free volume, these GB may also present dangling bonds and resemble structures found in amorphous graphene obtained by electron bombardment \cite{Meyer}.

The possibility of dangling bonds makes these high energy GB particularly interesting since the dangling bonds can carry magnetic moments and are potential sources of magnetic ordering \cite{YazyevREP}. The possibility that grain boundaries can be a source of magnetism in graphitic materials was suggested in Ref.\onlinecite{Cervenka} based on the following experimental observations in highly oriented pyrolytic graphite (HOPG). STM studies of  GB with different periodicities found some peaks in the local density of states attributed to dangling bonds. Depending on the periodicity of the GB, these additional peaks in the density of states were either situated at the Fermi energy or split, which was interpreted as spin splitting. The room temperature ferromagnetism measured by magnetic force microscopy and bulk magnetization measurements was tentatively attributed to two-dimensional magnetic ordering at the grain boundaries. The observation of room temperature ferromagnetism was, however,  not confirmed in other studies of HOPG \cite{PRLspanish}. Recently, a systematic study of samples of HOPG of different manufacturers \cite{Irina} has convincingly attributed the macroscopic magnetic signal found in some of them to Fe-rich inclusions buried in the bulk. Nevertheless, the  local STM data of Ref.\onlinecite{Cervenka} could still be related to the existence of localized magnetic moments and  the possibility to achieve ferromagnetism in $sp$ electron materials remains very  appealing \cite{Edwards2006} and justifies further research.

In this paper, we study systematically the structural, electronic and magnetic properties of GB with dangling bonds (DBGB) in graphene by a hierarchical approach based on classical atomistic simulations and ab-initio calculations. As a result of a massive search based on simulated annealing by classical Monte Carlo simulations, we find, that a particular structure with 5, 8 and 7 rings (5-8-7) appears to be kinetically stable up to high temperature and can be a common structural element of generic GB in graphene. According to our DFT calculations the 5-8-7 defect contains one dangling bond with an associated magnetic moment of $\simeq 0.5-1.0 \mu_B$ with $\mu_B$ the Bohr magneton, that is only partially reduced by hydrogenation. This means that, in contrast to the low energy GB, a generic GB in graphene and graphite can have unpaired electrons and magnetic moments. Note that, according to our calculations, the hydrogenation of DBGB turns out to be energetically favorable, thus, the most probably realistic DBGB in graphene should be passivated by hydrogen. At the same time, all qualitative conclusions about the structure and magnetism of GB do not depend on this assumption.

The paper is organized as follows. In section \ref{sec:method} we present the methods for atomistic simulations and ab-initio calculations. In section \ref{sec:energetics} we describe the structure and energetics of DBGB in graphene and in Section \ref{sec:electronics} we discuss their electronic structure and spin density. Finally, in Section \ref{sec:conclusions} we give a summary and conclusions.

\section{Method}
\label{sec:method}

A systematic study of GB is computationally demanding because it requires the examination of very large samples.
Therefore we have done a first search for DBGB by means of Monte Carlo simulations based on the classical LCBOPII interatomic potential \cite{LCBOPII}.
After having identified the 5-8-7 structure as a promising basic unit for DBGB, we have studied the electronic and magnetic properties by means of spin polarized
Density Functional Theory (DFT) calculations as implemented in the SIESTA code.
The drawing of flat pictures was done using the xyz2eps utility \cite{xyz2eps} written in Python Programming Language \cite{python}.
The visualization of 3D structures together with 3D charge density was done using the VESTA visualization program \cite{vesta}.

\subsection{Atomistic simulations with LCBOPII}

The classical bond-order potential LCBOPII \cite{LCBOPII} has been shown to describe accurately the structure \cite{NatMat,PRBripple} and elastic  properties \cite{PRL2009} of graphene as well as the phonons \cite{LendertJan}, the structure of the edges \cite{Jaap} and bilayer graphene \cite{bil}. The accuracy of this potential for dealing with GB has been validated against DFT calculations in Ref. \cite{Carlsson}. For the present study, this potential has the important feature of being reactive, namely to allow breaking and formation of bonds as it would happen when grains meet.

We have used Monte Carlo simulations in the NPT ensemble, namely we have kept temperature T and number of particles N constant and allowed volume fluctuations as to keep the pressure P=0. To find (meta)stable structures we have done a simulated annealing lowering the temperatures from 3300K. The procedure to construct the samples is described in section \ref{sec:gbmodel}.

\subsection{DFT ab-initio calculations with the SIESTA code}

    We have performed spin polarized DFT \cite{DFT-1, DFT-2} calculations
by means of the package SIESTA which implements DFT on a localized basis set
\cite{SIESTA-1, SIESTA-2, SIESTA-3}.
    We used GGA with Perdew-Burke-Ernzerhof parametrization (GGA-PBE)
\cite{GGA-PBE} and a standard built-in double-$\zeta$ polarized (DZP)
\cite{DZP-NAO} basis set to perform geometry relaxation of
graphene samples with GB.
    The DZP basis set represents core electrons by norm-conserving
Troullier-Martins pseudopotentials \cite{pseudopotentials} in the
Kleynman-Bylander nonlocal form \cite{nonlocal-form}. For a carbon atom
this basis set has 13 atomic orbitals: a double-$\zeta$ for 2s and 2p
valence orbitals and a single-$\zeta$ set of five d orbitals.
    The cutoff radii of the atomic orbitals were obtained from an
energy shift equal to 0.02 Ry which gives a cut-off radius of 2.22 \AA~
for s orbitals and 2.58 \AA~ for p orbitals.
    The real-space grid is equivalent to a plane-wave cutoff energy of 400 Ry,
yielding $\approx$ 0.08 \AA~ resolution for the sampling of real space.
    For non periodical directions, an extra space larger than 15 \AA~
was added to avoid spurious interactions.
    We used k-point sampling of the Brillouin zone based on the Monkhorst-Pack
scheme \cite{Monkhorst-Pack} where the number of k-points was defined similarly
to the k-grid cutoff radii equal to 15 \AA~ which usually gives 4-20 k-points
depending on the sample size.
    The geometries were relaxed using the conjugate gradient method until all
interatomic forces were smaller than 0.04 eV/\AA~ and the total stress
less than 0.0005 eV/\AA$^3$.
    No geometrical constrains were applied during relaxation.

\subsection{GB structural model}
\label{sec:gbmodel}

\begin{figure}
    \centering
    \includegraphics[width=0.4\textwidth]{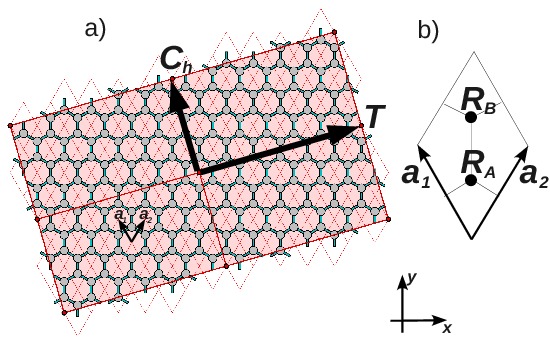}
    \caption{\label{fig:geomunitcell}
        (color online)
        Definition of chirality vector (a) and unit cell (b).
        For clarity the grain defined by vectors ${\bf C}_h$ and ${\bf T}$
        is replicated twice along the direction of ${\bf C}_h$ and ${\bf T}$}
\end{figure}

It is common practice to generate GB by means of the coincidence site lattice (CSL) and this theory has also been used to study low-energy GB in graphene \cite{Carlsson}.
The CSL theory, however, includes only symmetric grain configurations and is not suitable to deal with generic GB, like the DBGB we study here.
Therefore we use a more general model, inspired by the theory of nanotubes \cite{nanotubes} and similar to that used in Refs.~\onlinecite{Yazyev,Carlsson} for symmetric GB.

A nanotube is uniquely defined by a pair of integers $(m, n)$ relating the chirality vector ${\bf C}_h$ to the basis vectors
of the hexagonal lattice $({\bf a}_1, {\bf a}_2)$ as
\begin{equation}
{\bf C}_h=m {\bf a}_1 +n {\bf a}_2
\end{equation}
where
\begin{equation}
{\bf a}_1 = r_{cc} (-\sqrt{3}/2, 3/2) ~~~~~
{\bf a}_2 = r_{cc} (~\sqrt{3}/2, 3/2)
\end{equation}
and  $r_{cc}=1.42$ \AA~ is the interatomic distance in graphene giving $a_0= \sqrt{3} r_{cc}$ as lattice constant (see Fig. \ref{fig:geomunitcell}a).
For nanotubes, the vector ${\bf T}$ orthogonal to ${\bf C}_h$ gives the nanotube axis and ${\bf C}_h$ gives the direction of rolling. In terms of $(m, n)$ the vector ${\bf T}$ is given by
\begin{equation}
{\bf T} = \frac{t_1} {k} {\bf a}_1 + \frac{t_2} {k} {\bf a}_2
\end{equation}
where
\begin{equation}
t_1 = -m - 2 n ~~~~~ t_2 = 2 m + n
\end{equation}
and $k$ is the greatest common divisor of $|t_1|$ and $|t_2|$.

Furthermore we call ${\bf R}_A$ and ${\bf R}_B$ the positions of the two atoms in the unit cell of the hexagonal lattice. The case 
\begin{equation}
{\bf R}_A = r_{cc} (0, 1, 0) ~~~~~ {\bf R}_B = r_{cc} (0, 2, 0)
\end{equation}
is illustrated in the unit cell shown  in Fig.\ref{fig:geomunitcell}b.

While for nanotubes the vectors ${\bf C}_h$ and ${\bf T}$ are used to define a rectangle of given chirality to be rolled, for GB the chirality vector ${\bf C}_h$ determines the direction of the grain boundary while the rectangular area is the graphene grain, as shown in Fig.\ref{fig:geomunitcell}a.
In the CSL approach the second grain is symmetric with respect to the GB direction given by ${\bf C}_h$.

The length $d(m, n)$ of ${\bf C}_h$ in our basis is
\begin{equation}
d(m, n) = r_{cc} \sqrt{3 \Sigma}
\end{equation}
where
\begin{equation}
\Sigma = m^2 + m n + n^2
\label{Sigma}
\end{equation}

There may be different pairs $(m, n)$ that give the
same value of $\Sigma$. For example $\Sigma=91$ may be obtained by pairs
$(1, 9)$ and $(5, 6)$ so that, for $\Sigma=91$, Eq.\ref{Sigma}
has the 4 solutions $(1, 9); (9, 1); (5, 6); (6, 5)$.

The couples $(1, 9); (9, 1)$ and $(5, 6); (6, 5)$ are symmetric and
are described by a single tilt angle in the CSL theory whereas {\it e.g.}
the pair $(1, 9); (5, 6)$ is not symmetric and requires to define the two misorientation angles of the two grains

\begin{equation}
\cos {\phi_i} = \frac{2 m_i + n_i} {2 \sqrt{m_i^2 + m_i n_i + n_i^2}} ~~~ i=1,2
\end{equation}

In this way, by selecting two grains with the same $\Sigma$ we can satisfy
periodic boundary conditions also for non symmetric grains selecting
different chirality vectors ${\bf C}_{h1}$ and ${\bf C}_{h2}$ together with the orthogonal vectors ${\bf T}_1$ and ${\bf T}_2$.
This procedure allows to cut two rectangular grains with the same periodicity $d$ that, after
proper reorientation, can be joined together to form the GB, labeled now by two pairs of indexes $(m_1, n_1)$ and $(m_2, n_2)$. In case of symmetric grains i.e. $n_1 = n_2$ and $m_1 = m_2$ we can define $\theta = \phi_1 + \phi_2$.

Since the two grains are rectangular, the final structure
forms a rectangular unit cell which contains two grains with two GB.
This construction gives us a starting point for the search of metastable non symmetric GB that we describe in the next Section.

\subsection{Search of (meta)stable DBGB}
\label{sec:metasearch}

  Once the procedure for building GB considers also asymmetric grains,
most situations will yield structures with large strain and atoms that are too close to each other,
from 1.5 \AA~ till 0.1 \AA~ or even less. We have used two parameters to help the search for favorable structures.
First, we introduce the parameter $r_{min}$ which controls the minimal distance
between atoms. If two atoms are closer than $r_{min}$ then this pair is replaced
by a single atom with average coordinates. The parameter $r_{min}$
influences the density of atoms along the grain boundary.
  We have searched with different values, namely  $r_{min}=0.1,0.4,1.2$ \AA.
This procedure is physically justified because, in situation of crystal growth at high temperature, carbon atoms would be redistributed in such a way as to avoid too close overlap of the atomic cores.
The other free parameter in our scheme is the shift ${\bf r}_{sh}$ of the sublattice vectors
${\bf R}_A$ and ${\bf R}_B$
\begin{equation}
{\bf R}_A = {\bf r}_{sh} + r_{cc} (0, 0, 0) ~~~~~ {\bf R}_B = {\bf r}_{sh} + r_{cc} (0, 1, 0)
\end{equation}
We use two values
\begin{equation}
{\bf r}_{sh}^1=r_{cc}(0,0,0)~~~~~~{\bf r}_{sh}^2=r_{cc}(0,1,0)
\end{equation}
where ${\bf r}_{sh}^1$ puts the origin of the cell on one atom and ${\bf r}_{sh}^2$ gives the ${\bf R}_A$ and ${\bf R}_B$ shown in Fig. \ref{fig:geomunitcell}.

We use the freedom given by the procedure described above to construct thousands of initial configurations with GB.
For each configuration, we optimize the structure by annealing the sample from 3300K by Monte Carlo simulations in the NPT ensemble with the LCBOPII interatomic potential. After a large number of Monte Carlo moves, we find structures that do not evolve anymore and can be considered as metastable.
Among all these configurations we search automatically the ones with two-fold coordinated carbon atoms.

Among these possibilities, the structure with 5-8-7 rings (see Fig.~\ref{fig:57-587-H587}) is the simplest and most common.
Therefore we have concentrated on this structure as prototype of DBGB.
For simplicity, we have then constructed samples with 5-7-8 DBGB and different periods with symmetric grains defined by $(m, n)$ and $\theta$.
Further relaxation of the selected structure with SIESTA affects the structure of graphene GB only marginally, which confirms the accuracy of our atomistic energy minimization.
Lastly we calculate electronic and magnetic properties with SIESTA.

\begin{figure}
    \centering
    \includegraphics[width=0.45\textwidth]{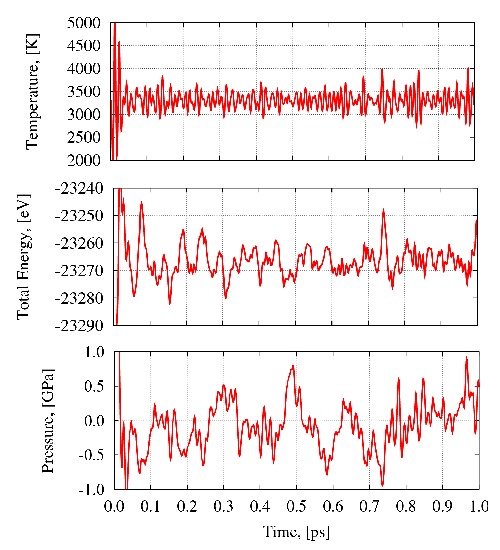}
    \caption{\label{fig:MD}
        (color online)
        Time dependence of temperature $T$, total energy $E_T$
        and pressure (total stress) $P$ during 1000 steps of MD annealing
        for a sample with GB with two 5-8-7 defects and 150 atoms.
        The mean values are
        $M(T)=3299.39$ K, $M(E_T)=-23266.89$ eV, $M(P)=-0.056$ GPa
        with standard deviation
        $\sigma(T)=187.64$ K, $\sigma(E_T)=4.80$ eV, $\sigma(P)=0.327$ GPa
        and correlation
        $\rho(T, E_T)=0.039$, $\rho(P, E_T)=-0.247$, $\rho(T, P)=0.177$.
    }
    %
\end{figure}

We have checked the stability of the 5-8-7 DBGB also by performing constant-temperature
Molecular Dynamics (MD) with Nose thermostat using the DFT package SIESTA at 3300K with a time step of 1 fs.
The time dependence of temperature, energy and pressure are shown in Fig.~\ref{fig:MD}.
After 1000 MD steps the structure of the 5-8-7 defect keeps its original geometry.
During the dynamics, however, we observe an exchange of a 6-ring with a 7-ring that causes a mirror reflection of the 5-8-7 point defect with respect to the GB line.
This transformation that keeps the original structure of the two-fold coordinated atom is shown in Fig.~\ref{fig:57to587}d.

\section{Structure and energetics of DBGB in graphene}
\label{sec:energetics}

\begin{figure}
    \centering
    \includegraphics[width=0.5\textwidth]{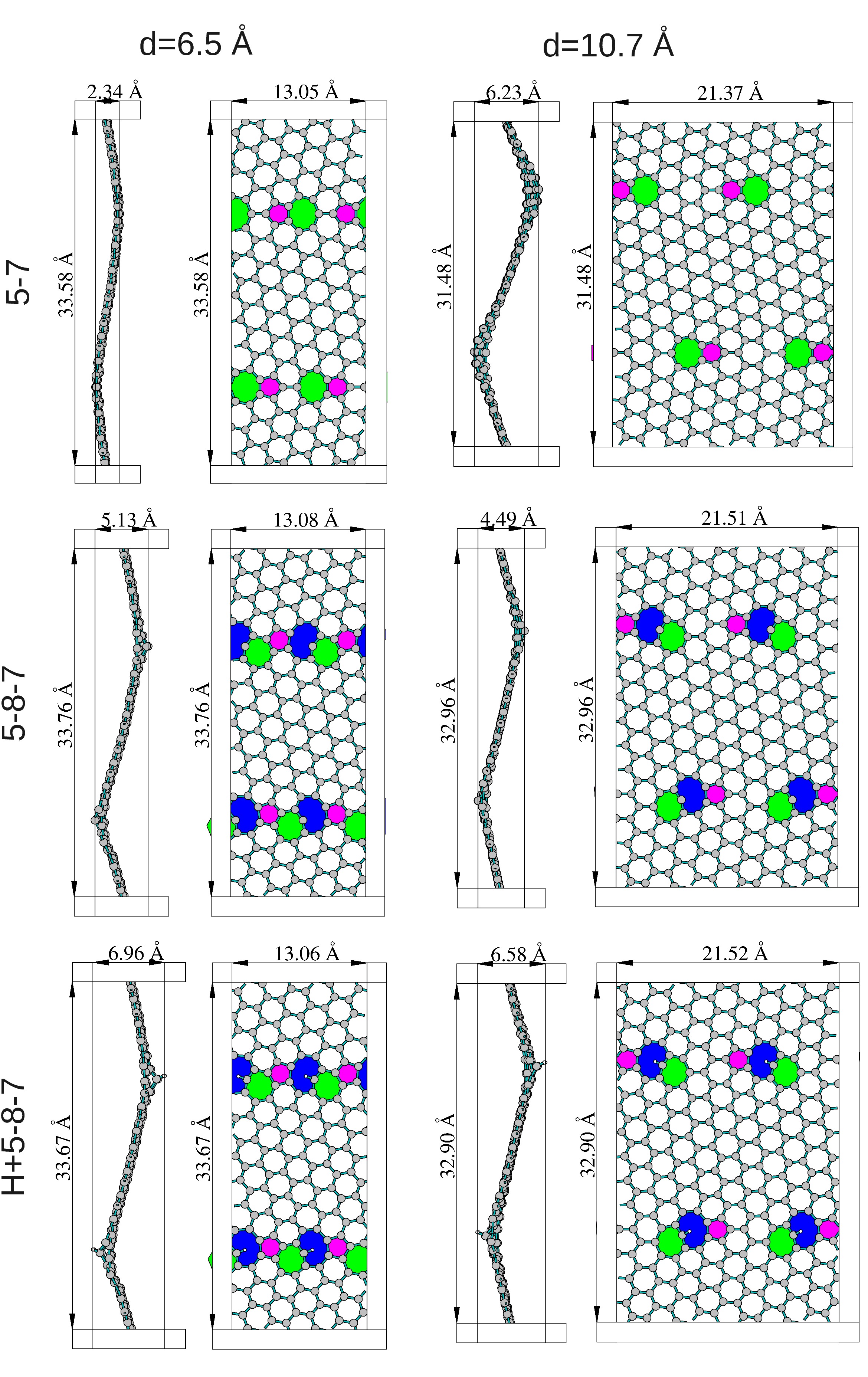}
    \caption{\label{fig:57-587-H587}
        (color online)
        Side and top view of (from top to bottom) 5-7, 5-8-7, H+5-8-7 GB for
        two values of the period $d$, left: $d=6.5$ \AA, right: $d=10.7$ \AA.
        The unit cell is replicated twice in the GB direction.
        For clarity, 7-rings are green (light gray),
        5-rings are pink (gray) and 8-rings are blue (black). 
    }
\end{figure}

\begin{figure}
    \centering
    \includegraphics[width=0.5\textwidth]{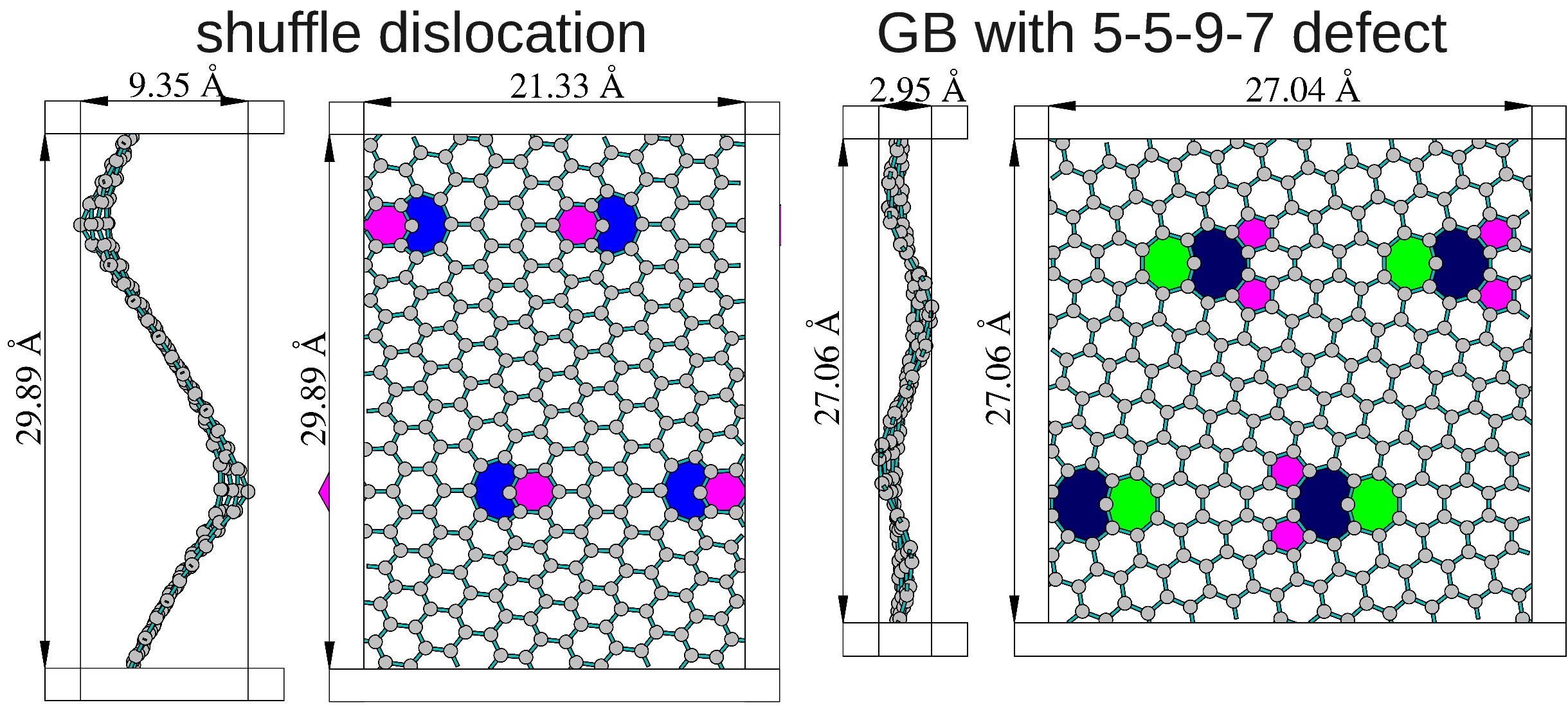}
    \caption{\label{fig:shuffle-and-5597}
        (color online)
        Side and top view of shuffle (left) and 5-5-9-7 GB. The unit cell is replicated twice in the GB direction.
        For clarity, in the shuffle GB we color also the 6-ring.
    }
\end{figure}

A more general way of describing GB is to present them as arrays of dislocations \cite{HirthLother}.
Low energy symmetric GB are nothing but arrays of 5-7 (glide) dislocations.
The DBGB that we select in our search for metastable structures  contain more complicated structural elements characterized by the presence of 8-, 9- and 4-fold rings.
These rings appear also in simulations of disordered graphene \cite{Malola} and graphene at high temperature \cite{JCM2011},
and were experimentally observed in electron bombarded graphene \cite{Meyer}.

In Ref.~\cite{VozmedPRB} another type of dislocation, the shuffle dislocation shown in
Fig.~\ref{fig:shuffle-and-5597}, with one 8-fold ring with one dangling bond, has been proposed as a potential carrier of a magnetic moment.
In our search for metastable structures with dangling bonds, we have found 8-fold rings only in combination with other non-hexagonal rings.
If we construct a 8-ring shuffle dislocation we find that above 2400K it transforms to a 5-8-7 configuration (these two dislocation configurations are characterized by the same Burgers vector as will be discussed in detail below). By looking at Figs.~\ref{fig:57-587-H587} and \ref{fig:shuffle-and-5597} one can see that the shuffle GB (i.e. the wall of shuffle dislocations) has the largest out-of-plane distortion,  which increases the strain in the structure \cite{Carlsson} and might explain its instability.

\begin{figure}
    \centering
    \includegraphics[width=0.45\textwidth]{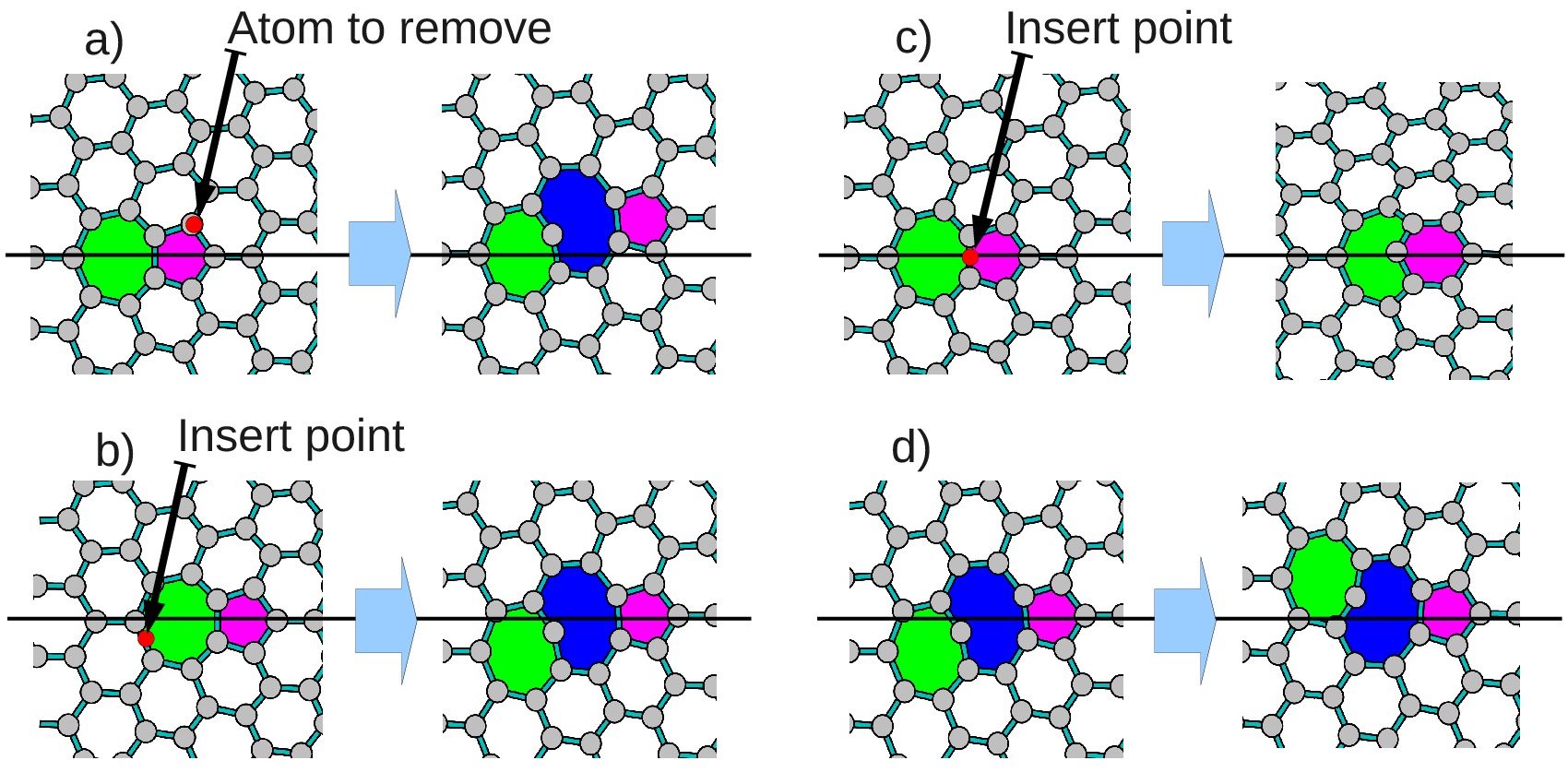}
    \caption{\label{fig:57to587}
        (color online)
        The 5-7 and 5-8-7 defects are related geometrically.
        There are two ways to construct a 5-8-7 defect from 5-7:
        remove an atom from pentagon (a),
        insert adatom to the bond belonging to heptagon (b).
        In the same way we can construct a shuffle dislocation from 5-7 (c).
        In panel d) we show the mirror transformation of the 5-8-7
        observed in the MD simulations at T=3300 K described in
        section~\ref{sec:metasearch}.
    }
\end{figure}

In our Monte Carlo simulations at 3300K, we find most frequently the sequence 5-8-7 which has one two-fold coordinated carbon atom.
This atom has one unpaired electron and, as a result, is the source of
magnetic moment. We call this atom therefore a magnetic atom.
If we remove the magnetic atom and apply further relaxation
we find the non magnetic 5-7 defect. In Fig.~\ref{fig:57to587} we show how the 5-8-7 is related to the 5-7 defect and how it can be constructed by either adding (Fig.~\ref{fig:57to587}b) or removing (Fig.~\ref{fig:57to587}a) an atom from it. The similar construction of a shuffle defect is shown in Fig.~\ref{fig:57to587}c.
This procedure is technically reversible so that a 5-7 can be obtained by removing the magnetic atom and letting the structure rebound and relax.

\begin{table}[ht]
    \caption{Summary of the studied defects with
             GB period $d$,
             Burgers vector ${\bf b}$,
             GB formation energy $E_F$
             and hydrogen adsorption energy (with respect to the hydrogen atom) $E_{ads}$.
             The tilt angle together with Burgers vector were calculated for
             z-projected geometries i.e. completely flat samples with z=0.
             The binding energy of the hydrogen molecule in the used model is $E_{H_2}$ = 4.53 eV.
             }
    \centering 
    \begin{tabular}{c c c c c} 
        \hline\hline 
        GB        & GB period & Tilt angle       & Burgers vector & $E_F$              \\
                  & $d$ (\AA) & $\theta^{\circ}$ & ${\bf b}$ (\AA) & (eV/defect)        \\
        \hline
        5-7       &   6.52    & 20.8             & 2.360           &  2.31              \\ 
        5-8-7     &   6.54    & 21.7             & 2.467           &  6.83              \\
        H+5-8-7   &   6.53    & 21.7             & 2.461           &  $E_{ads}$ = 4.78  \\
        \hline
        5-7       &  10.69    & 13.7             & 2.544           &  3.87              \\ 
        5-8-7     &  10.76    & 12.7             & 2.378           &  8.01              \\
        H+5-8-7   &  10.76    & 12.5             & 2.350           &  $E_{ads}$ = 4.60  \\
        shuffle   &  10.66    & 13.2             & 2.451           &  8.16              \\
        \hline
        5-5-9-7   &  13.59    & 17.7             & 4.185           &  8.63              \\
        \hline 
    \end{tabular}
    \label{tab:defects} 
\end{table}

\begin{figure}
    \centering
    \includegraphics[width=0.45\textwidth]{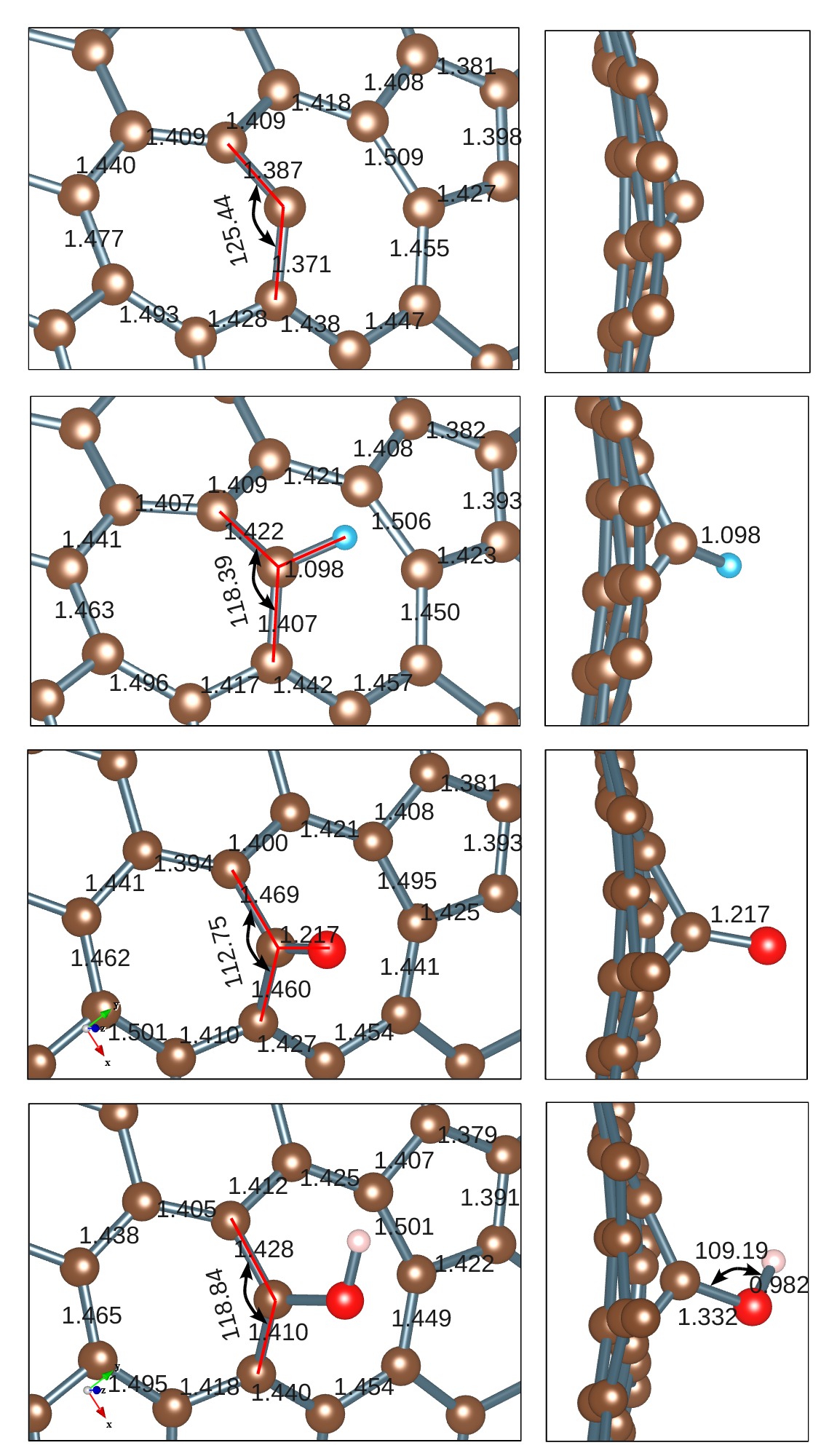}
    \caption{\label{fig:core-587-587H-587O-587OH}
        (color online) Bond lengths and C-C-C angle of magnetic atom
        for (from top to bottom) 5-8-7, H+5-8-7, O+5-8-7, OH+5-8-7 DBGB.
    }
\end{figure}

One could expect the 5-8-7 DBGB to have the same  Burgers vector  of the glide and shuffle dislocation. In fact,
if we consider the dislocation as a disclination dipole \cite{HirthLother} the Burgers vector ${\bf b}$ is the product of the Frank vector of the disclination times the dipole arm.
If we double the distance between the 5 and 7-fold rings that constitute the disclination,
we could expect a twice larger Burgers vector ${\bf b} \rightarrow 2{\bf b}$.
The 8-fold ring between the 5- and 7-fold rings can be considered as a shuffle dislocations
with Burgers vector $- {\bf b}$ so that the resulting Burgers vector is $2{\bf b} -{\bf b} ={\bf b}$.

This analysis is supported by the data shown in Table~\ref{tab:defects} where
we compare GB made of arrays of the 5-7 and 5-8-7 disclination dipoles shown in Fig.~\ref{fig:57-587-H587}.
The Burgers vector was calculated using the Frank equation \cite{HirthLother}
\begin{equation}
b=2 d \sin{\theta/2}
\end{equation}
where $d$ is the periodicity of the array and $\theta$ is the misorientation angle.
One can indeed see that the Burgers vector of the 5-7 and 5-8-7 are almost the same.
We also compare the formation energy of defects $E_F$ calculated as
\begin{equation}
E_F = (E_{Total}^{Defect} - E_{Total}^{Graphene} \frac{N_{atoms}^{Defect}}{N_{atoms}^{Graphene}})/ N_{Defects}
\end{equation}
for different types of GB. The formation energy of the 5-8-7 GB is approximately twice the one of the 5-7 for the same periodicity, which is not surprising since the dangling bond costs some additional energy. The larger formation energy for the larger periodicity is consistent with the finding~\cite{Carlsson} that dislocation cores attract each other, contrary to three-dimensional materials. 

The presence of the dangling bond makes bonding to other species possible.
We have therefore studied the 5-8-7 also when the magnetic atom is bound to a hydrogen atom, a structure we call H+5-8-7, or to an oxygen atom or OH group, that we call O+5-8-7 and  OH+5-8-7 respectively .
The top and side view of H+5-8-7 shown in Fig.~\ref{fig:57-587-H587} do not differ much from the 5-8-7.
Only the local structure of the magnetic atom is somewhat changed.
In particular, the bonds to its two carbon neighbors go from $\sim 1.37 $ \AA~ in 5-8-7 to $\sim 1.41 $ \AA~ in H+5-8-7,
a value closer to the bulk value 1.42 \AA. The angle between these two bonds is also changed. The rest of the structure remains basically the same
as shown in Fig. ~\ref{fig:core-587-587H-587O-587OH} also for the case of oxygen and OH.

It is remarkable that the adsorption energy of the H+5-8-7 is just a bit higher than the H$_2$ binding energy calculated within the same method. This means that, within our computational scheme, the hydrogenation of DBGB is energetically favorable. At the same time, the difference is small and one should take into account that the density functional within GGA underestimates strongly the binding energy of $H_2$ molecule. Fortunately, the issue of the hydrogenation does not affect qualitatively our conclusions about the structure (as it is shown here) and magnetism (as will be shown below) of DBGB.

Since the 5-8-7 DBGB has minimal Burgers vector and low strain in view of its flatness it is natural to assume that it has the lowest energy among DBGB and therefore represents the most natural candidate as source of magnetism
in GB. That is why we will focus on this structural element in the rest of our paper. Of course more complicated DBGB exist and, as an example, we show in Fig.\ref{fig:shuffle-and-5597}
the structure of a GB with $\theta=17.7^\circ$ formed by a periodic array of a 5-5-9-7
structural element. As reported in Table~\ref{tab:defects}, this GB has formation energy  just slightly higher than the 5-8-7 and an almost double Burgers vector. The latter statement is justified by taking into account the change of type of GB from zigzag to armchair \cite{Yazyev}. In the following section we examine in detail the 5-8-7 in comparison to the 5-7.

\section{Electronic structure and spin density of DBGB in graphene}
\label{sec:electronics}

\begin{figure}
    \centering
    \includegraphics[width=0.45\textwidth]{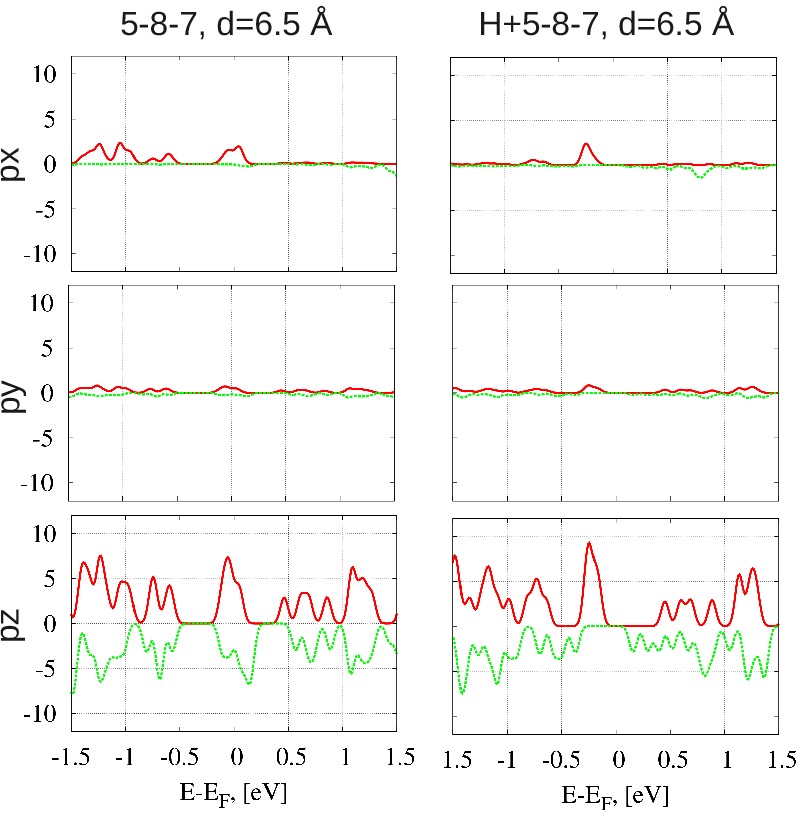}
    \caption{\label{fig:dos-H-587}
        (color online) p$_x$, p$_y$ and p$_z$ components of total DOS for 
       5-8-7 and H+5-8-7 DBGB with misorientation angle $\theta=21.7^\circ$. Red solid and dashed green curves are for spin-up and spin-down respectively.}
\end{figure}

\begin{figure}
    \centering
    \includegraphics[width=0.45\textwidth]{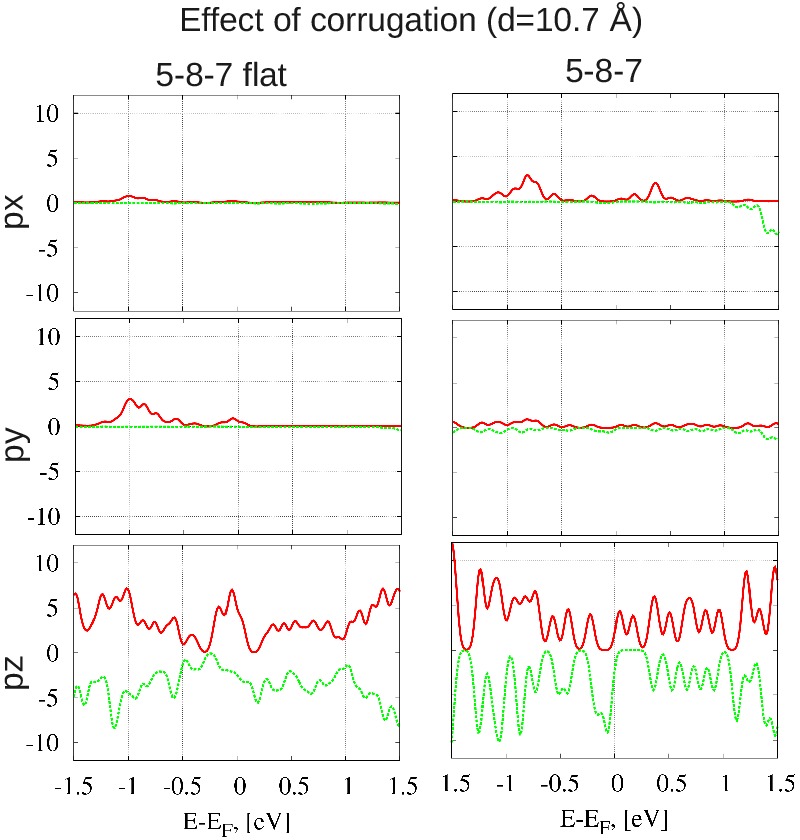}
    \caption{\label{fig:dos-587-corrugation}
        (color online)
        The effect of corrugation on the DOS (see the text) is seen by comparing the minimal energy 
        5-8-7 DBGB with the flat one. Red solid and dashed green curves are for spin-up and spin-down respectively.}
\end{figure}

\begin{figure}
    \centering
    \includegraphics[width=0.42\textwidth]{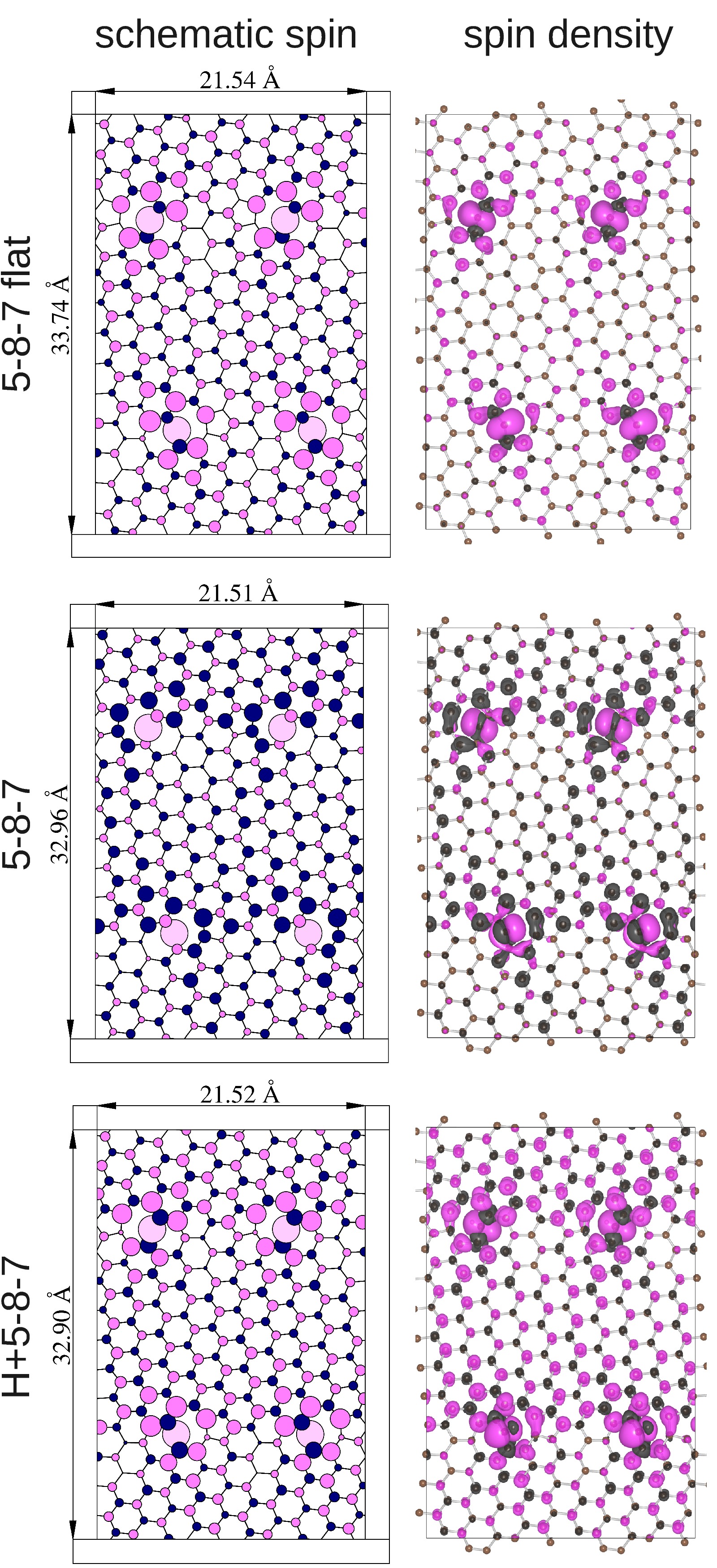}
    \caption{\label{fig:587-spin}
        (color online)
        Right panels: isosurface plot of the spin density for the 5-8-7 DBGB with $\theta=12.7^\circ$. 
        Left panels: symbolic representation of the spins per atom (see text).
        Pink (gray) and blue (dark gray) represent spin-up and spin-down, respectively. In the left panels the magnetic atoms are light pink (light gray).
        From top to bottom: flat configuration; minimum energy configuration without hydrogen; 
        minimum energy configuration with hydrogen.
    }
\end{figure}

\begin{figure}
    \centering
    \includegraphics[width=0.42\textwidth]{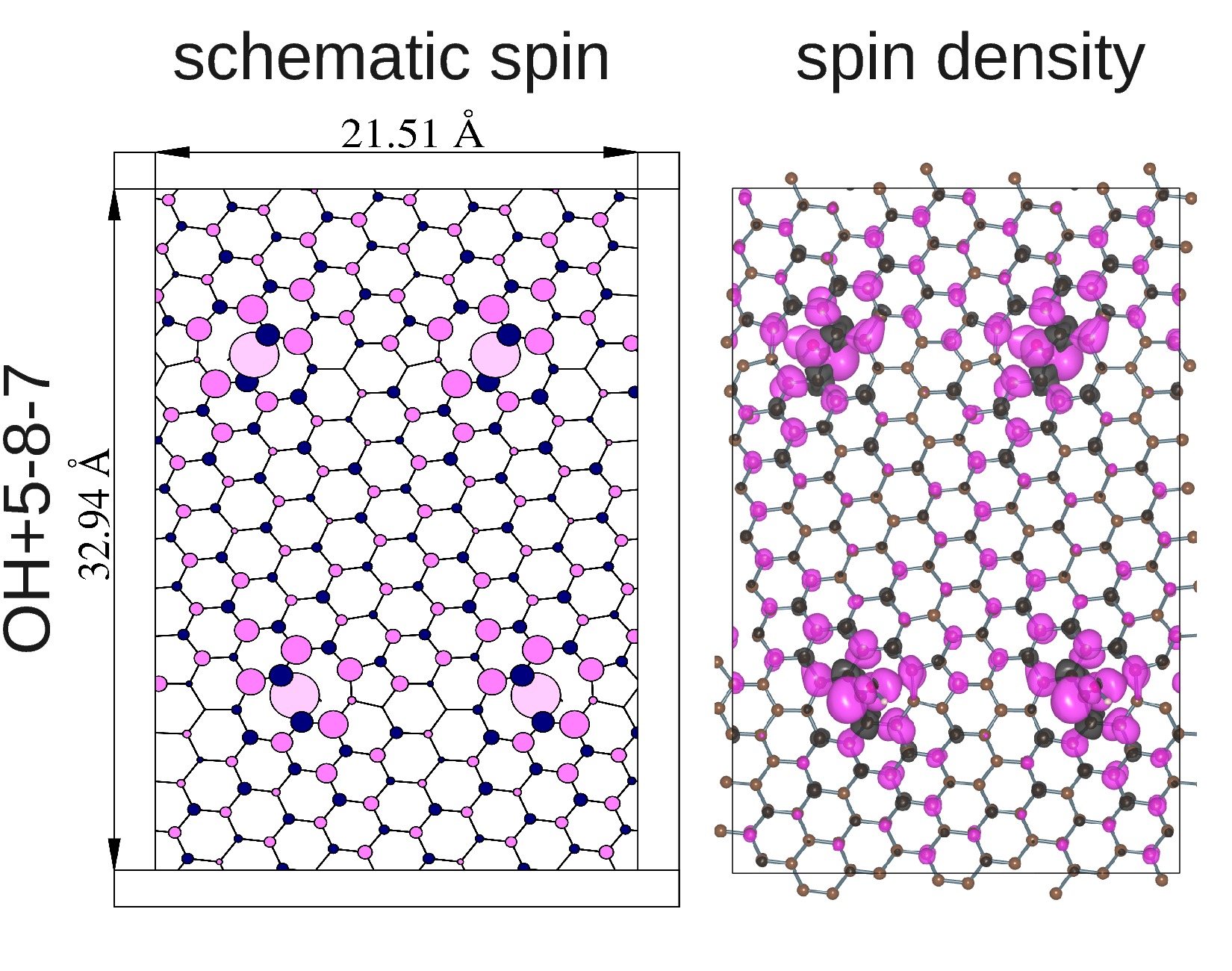}
    \caption{\label{fig:587OH-spin}
        (color online)
        Right panel: isosurface plot of the spin density for the OH+5-8-7 DBGB with $\theta=12.7^\circ$. 
        Left panel: symbolic representation of the spins per atom (see text).
        Pink (gray) and blue (dark gray) represent spin-up and spin-down, respectively.
        In the left panels the magnetic atoms are light pink (light gray).
    }
\end{figure}

\begin{table}[ht]
    \caption{Magnetic moment contribution in $\mu_B$ from magnetic atom
        which carries dangling bond (DB atom)
        for the studied samples with two GB and one magnetic defect per GB.}
    \centering
    \begin{tabular}{c c c c c}
        \hline
        \hline
                                      & from     & number & from   & from   \\
        System                        & DB       & of DB  & non DB & whole  \\
                                      & atom     & atoms  & atoms  & system \\
        \hline
        {H+5-8-7 $d$=6.5 (\AA)}       & 0.530    & 2      &  0.956 & 2.001 \\
        {5-8-7 $d$=6.5 (\AA)}         & 0.933    & 2      &  0.843 & 2.708 \\
        {H+5-8-7 flat $d$=6.5 (\AA)}  & 0.496    & 2      &  1.019 & 1.997 \\
        {5-8-7 flat $d$=6.5 (\AA)}    & 0.921    & 2      &  0.850 & 2.692 \\
        \hline
        {OH+5-8-7 $d$=10.7 (\AA)}     & 0.291    & 2      &  0.468 & 1.068 \\
        {H+5-8-7 $d$=10.7 (\AA)}      & 0.542    & 2      &  0.895 & 1.963 \\
        {5-8-7 $d$=10.7 (\AA)}        & 0.455    & 2      & -0.742 & 0.169 \\
        {H+5-8-7 flat $d$=10.7 (\AA)} & 0.455    & 2      &  1.037 & 1.933 \\
        {5-8-7 flat $d$=10.7 (\AA)}   & 0.987    & 2      &  0.990 & 2.964 \\
        \hline
        {shuffle $d$=10.7 (\AA)}      & 0.424    & 2      & -0.011 & 0.837 \\
        {5-5-9-7 $d$=13.6 (\AA)}      & 0.889    & 2      & -0.333 & 1.445 \\
        \hline
    \end{tabular}
    \label{tab:spin}
\end{table}


We have calculated the spin polarized density of states (DOS)  of selected DBGB by means of SIESTA.
We project the states onto the orbitals representing the p$_x$, p$_y$ and p$_z$.
In Appendix A we validate our approach against previous results for
the H-saturated zigzag graphene edges~\cite{YazyevEdgeMagnetism, EdgeHydrogen} while comparing them to non saturated edges.
In Fig.\ref{fig:dos-H-587} we present the spin polarized DOS for
ferromagnetically oriented magnetic moments associated to the dangling bond of a 5-8-7 and H+5-8-7 with period $d=6.5$ \AA.
We see that DOS is mostly p$_z$ and is essentially different for spin up and spin down.
For the H+5-8-7 there is even  an almost half-metallic situation with the Fermi energy lying just below the gap for majority spin electron states.
Below the Fermi energy but relatively far from it, there is also a smaller gap for minority electron states.
The tiny p$_x$, p$_y$ components are related to the distortion from a planar sp$^2$ bond.
In Table~\ref{tab:spin} we report the magnetic moments per magnetic atom.
They are in general not integer.
Importantly, hydrogen adsorption does not destroy the magnetic moment.
This is because the magnetic atom is not like a usual dangling bond that can be fully saturated by hydrogen.
A carbon atom participates with three electrons to in-plane bonding and with the fourth to the p$_z$ band.
Therefore the two-fold coordination in the plane provides a dangling bond that adds to and distorts the p$_z$ orbital.
The OH group reduces further the magnetic moment whereas oxygen destroys it completely.  

Lastly, we have found that the out of plane corrugation affects the magnetic moment of the 5-8-7 while it is not important for the H+5-8-7. In principle this effect can be used to  control magnetic moments through strain and therefore it deserves a more detailed discussion. To this aim, in Fig.~\ref{fig:dos-587-corrugation} we compare the DOS of the 5-8-7 with the one obtained for the same structure without allowing out-of-plane distortions,
namely for a flat 5-8-7. Since the out of plane corrugation is larger for $d=10.7$\AA~ (see, Fig.\ref{fig:57-587-H587})
we have chosen this case to illustrate this effect. One can see that the DOS are essentially different for the cases with and without out of plane deformations. The different DOS are also reflected in the almost double value of the magnetic moments of the flat 5-8-7 as reported in Table~\ref{tab:spin}. Conversely, the magnetic moments of relaxed and flat H+5-8-7 are comparable.

To understand the origin of this effect we have studied the spin density in the system.
In Fig.\ref{fig:587-spin} and Fig.\ref{fig:587OH-spin} we use two representations of the spin density.
The one to the right is the most common representation of isosurfaces of the spin density.
The representation to the left, gives the amount of spin per atom obtained from Mullikan population analysis
represented as a sphere of radius proportional to the logarithm of the spin.
This representation makes it possible to visualize also the small spin density components.
In this way one can see that the up and down components away from the defect seem to be located on the A and B sublattices of graphene.
This alternation is broken by the defect in a way that depends on the out of plane distortions.
In fact, in the flat 5-8-7, the magnetic atom (light gray) with spin up has the two nearest neighbor of spin down
whereas in the relaxed 5-8-7 the nearest neighbors have the same spin up of the magnetic atom.

The H+5-8-7 is not sensitive to the corrugation and the spin distribution for
the flat case is very similar to the one shown for the relaxed H+587 in Fig.\ref{fig:587-spin}.

\begin{table}[ht]
    \caption{Magnetic moment contribution in $\mu_B$
             from A- and B-sublattices for 4 studied cases.}
    \centering
    \begin{tabular}{ c c c c c c }
        \hline
        \hline
            system &  $m_A$ &  $m_B$ &  $m_A+m_B$ & with H & $m_A+m_B$ \\
        \hline
            zz     &  1.464 & -0.154 &  1.310     & -      & 1.29 from \cite{EdgeHydrogen}\\
            H+zz   &  0.453 & -0.134 &  0.330     & 0.310  & 0.30 from \cite{YazyevEdgeMagnetism}\\
            2H+zz  & -0.233 &  0.738 &  0.505     & 0.625  & -      \\
            C+zz   & -0.080 &  0.390 &  0.310     & -      & -      \\
        \hline
    \end{tabular}
    \label{tab:zz-test-magnetic-moment} 
\end{table}

\begin{figure}
    \centering
    \includegraphics[width=0.45\textwidth]{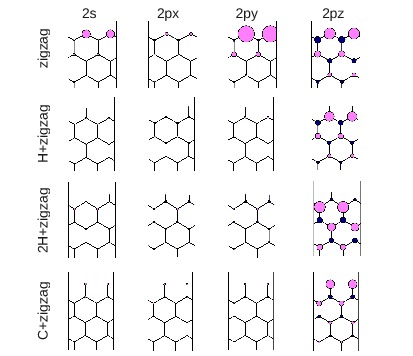}
    \caption{\label{fig:zz-test}
        (color online)
        Zigzag, single hydrogenated zigzag, double hydrogenated zigzag and
        model zigzag graphene edge with carbon. For each case the
        magnetic moment decomposition over atomic orbitals for carbon atoms only is shown according to
        Mulliken atomic orbital population analysis. The square of each circle is proportional to the
        value of magnetic moment contribution. Pink (light gray) and blue (dark gray) are
        positive and negative values of spin respectively.
        }
\end{figure}

\begin{table}[ht]
    \caption{Detailed information about distribution of
        magnetic moment over orbitals for 1-fold (1nn) 2-fold (2nn) 3-fold (3nn) coordinated edge carbon atoms
        in the four different systems shown in Fig.\ref{fig:zz-test}, i.e. zz for zigzag,
        H+zz for single hydrogenated zigzag, 2H+zz for double hydrogenated zigzag and
        C+zz for model zigzag graphene edge with carbon atom at the edge}
    \centering
    \begin{tabular}{ c c c c c c c }
        \hline
        \hline
            C-atom  & 2s & 2px & 2py & 2pz & all-d & sum \\
        \hline
        {zz} 2nn    &  0.176 &  0.026 &  0.691 &  0.320 & -0.009 &  1.206 \\
        {zz} 3nn    &  0.002 & -0.010 & -0.012 & -0.112 &  0.075 & -0.060 \\
        \hline
        {H+zz} 2nn  &  0.001 &  0.007 &  0.010 &  0.266 & -0.002 &  0.290 \\
        {H+zz} 3nn  &  0.002 & -0.008 & -0.006 & -0.071 &  0.025 & -0.057 \\
        \hline
        {2H+zz} 2nn & -0.004 & -0.013 & -0.015 & -0.019 &  0.036 & -0.017 \\
        {2H+zz} 3nn &  0.012 &  0.008 &  0.010 &  0.345 &  0.000 &  0.375 \\
        \hline
        {C+zz} 1nn  &  0.010 &  0.011 &  0.009 &  0.226 &  0.002 &  0.256 \\
        {C+zz} 2nn  &  0.004 & -0.004 & -0.007 & -0.048 &  0.018 & -0.038 \\
        {C+zz} 3nn  &  0.002 &  0.002 &  0.000 &  0.074 & -0.003 &  0.075 \\
        \hline
    \end{tabular}
    \label{tab:zz-test} 
\end{table}

\section{Conclusions}
\label{sec:conclusions}

Grain boundaries (GB) seem to be unavoidable structural elements of large enough graphene samples, irrespective of their preparation. 
By analogy with conventional three-dimensional material science, one may expect that they will affect strongly both the mechanical and electronic properties of graphene.
From a theoretical point of view, GB are very complicated objects due to the multiscale character of the problem.
Both long-range deformations extending over tens of thousands of atoms and specific atomistic and electronic structure of the cores are essential.
Therefore usually people study only special GB, mostly, those which can be constructed by the CSL approach \cite{Carlsson,Yazyev}. 
These GB are, indeed, usually the most energetically favorable. 
At the same time, e.g. for CVD growth of graphene on metals \cite{LiScience,Bae}, one could expect that various crystallites grow independently from many centers and more complicated GB will be formed. 
To attack this problem we have combined large-scale atomistic simulations using the LCBOBII potential \cite{LCBOPII} with ab-initio calculations. 
We have studied in detail GB containing the 5-8-7 defect which is the carrier of magnetic moment. 
Based on the results presented here one can conclude that a generic GB should contain magnetic moments which are robust enough, in particular, with respect to hydrogenation. 
Since GB in graphene are one-dimensional objects, they  cannot lead to magnetic ordering at any finite temperature. We have shown, however, that the very existence of magnetic moments at the GB dangling bonds modifies the local electronic structure around the Fermi energy that can be probed by STM.

\section{Acknowledgments}

This work is part of the research program of the "Stichting voor Fundamenteel Onderzoek der Materie (FOM)," 
which is financially supported by the “Nederlandse Organisatie voor Wetenschappelijk Onderzoek (NWO).”
The computational resources were provided by the Netherlands National Computing Facilities foundation 
(Stichting National Computerfaciliteiten, NCF) on SARA supercomputer facility at Amsterdam.

~

\appendix
\section{Validation test for magnetism of zigzag graphene edge with and without hydrogenation}
\label{sec:validation}

To check our computational scheme we have carried out electronic structure calculations for two cases 
where graphene is supposed to be magnetic, namely, zigzag edges \cite{YazyevEdgeMagnetism,EdgeHydrogen} 
with and without passivation by single- and double hydrogen and carbon-terminated zigzag edge (see Fig. \ref{fig:zz-test}). 
The results are shown in  Table \ref{tab:zz-test-magnetic-moment}.
One can see that in all cases we have an excellent agreement with previous results. Furthermore, 
we present in Table \ref{tab:zz-test} a more detailed information about the orbital contributions to the magnetic moments.


\begin{thebibliography}{52}%
\makeatletter
\providecommand \@ifxundefined [1]{%
 \@ifx{#1\undefined}
}%
\providecommand \@ifnum [1]{%
 \ifnum #1\expandafter \@firstoftwo
 \else \expandafter \@secondoftwo
 \fi
}%
\providecommand \@ifx [1]{%
 \ifx #1\expandafter \@firstoftwo
 \else \expandafter \@secondoftwo
 \fi
}%
\providecommand \natexlab [1]{#1}%
\providecommand \enquote  [1]{``#1''}%
\providecommand \bibnamefont  [1]{#1}%
\providecommand \bibfnamefont [1]{#1}%
\providecommand \citenamefont [1]{#1}%
\providecommand \href@noop [0]{\@secondoftwo}%
\providecommand \href [0]{\begingroup \@sanitize@url \@href}%
\providecommand \@href[1]{\@@startlink{#1}\@@href}%
\providecommand \@@href[1]{\endgroup#1\@@endlink}%
\providecommand \@sanitize@url [0]{\catcode `\\12\catcode `\$12\catcode
  `\&12\catcode `\#12\catcode `\^12\catcode `\_12\catcode `\%12\relax}%
\providecommand \@@startlink[1]{}%
\providecommand \@@endlink[0]{}%
\providecommand \url  [0]{\begingroup\@sanitize@url \@url }%
\providecommand \@url [1]{\endgroup\@href {#1}{\urlprefix }}%
\providecommand \urlprefix  [0]{URL }%
\providecommand \Eprint [0]{\href }%
\providecommand \doibase [0]{http://dx.doi.org/}%
\providecommand \selectlanguage [0]{\@gobble}%
\providecommand \bibinfo  [0]{\@secondoftwo}%
\providecommand \bibfield  [0]{\@secondoftwo}%
\providecommand \translation [1]{[#1]}%
\providecommand \BibitemOpen [0]{}%
\providecommand \bibitemStop [0]{}%
\providecommand \bibitemNoStop [0]{.\EOS\space}%
\providecommand \EOS [0]{\spacefactor3000\relax}%
\providecommand \BibitemShut  [1]{\csname bibitem#1\endcsname}%
\let\auto@bib@innerbib\@empty
\bibitem [{\citenamefont {Berger}\ \emph {et~al.}(2004)\citenamefont {Berger},
  \citenamefont {Song}, \citenamefont {Li}, \citenamefont {Li}, \citenamefont
  {Ogbazghi}, \citenamefont {Feng}, \citenamefont {Dai}, \citenamefont
  {Marchenkov}, \citenamefont {Conrad}, \citenamefont {First},\ and\
  \citenamefont {de~Heer}}]{Berger1}%
  \BibitemOpen
  \bibfield  {author} {\bibinfo {author} {\bibfnamefont {C.}~\bibnamefont
  {Berger}}, \bibinfo {author} {\bibfnamefont {Z.}~\bibnamefont {Song}},
  \bibinfo {author} {\bibfnamefont {T.}~\bibnamefont {Li}}, \bibinfo {author}
  {\bibfnamefont {X.}~\bibnamefont {Li}}, \bibinfo {author} {\bibfnamefont
  {A.~Y.}\ \bibnamefont {Ogbazghi}}, \bibinfo {author} {\bibfnamefont
  {R.}~\bibnamefont {Feng}}, \bibinfo {author} {\bibfnamefont {Z.}~\bibnamefont
  {Dai}}, \bibinfo {author} {\bibfnamefont {A.~N.}\ \bibnamefont {Marchenkov}},
  \bibinfo {author} {\bibfnamefont {E.~H.}\ \bibnamefont {Conrad}}, \bibinfo
  {author} {\bibfnamefont {P.~N.}\ \bibnamefont {First}}, \ and\ \bibinfo
  {author} {\bibfnamefont {W.~A.}\ \bibnamefont {de~Heer}},\ }\href {\doibase
  10.1021/jp040650f} {\bibfield  {journal} {\bibinfo  {journal} {J. Phys. Chem.
  B}\ }\textbf {\bibinfo {volume} {108}},\ \bibinfo {pages} {19912} (\bibinfo
  {year} {2004})}\BibitemShut {NoStop}%
\bibitem [{\citenamefont {Berger}\ \emph {et~al.}(2006)\citenamefont {Berger},
  \citenamefont {Song}, \citenamefont {Li}, \citenamefont {Wu}, \citenamefont
  {Brown}, \citenamefont {Naud}, \citenamefont {Mayou}, \citenamefont {Li},
  \citenamefont {Hass}, \citenamefont {Marchenkov}, \citenamefont {Conrad},
  \citenamefont {First},\ and\ \citenamefont {de~Heer}}]{Berger2}%
  \BibitemOpen
  \bibfield  {author} {\bibinfo {author} {\bibfnamefont {C.}~\bibnamefont
  {Berger}}, \bibinfo {author} {\bibfnamefont {Z.}~\bibnamefont {Song}},
  \bibinfo {author} {\bibfnamefont {X.}~\bibnamefont {Li}}, \bibinfo {author}
  {\bibfnamefont {X.}~\bibnamefont {Wu}}, \bibinfo {author} {\bibfnamefont
  {N.}~\bibnamefont {Brown}}, \bibinfo {author} {\bibfnamefont
  {C.}~\bibnamefont {Naud}}, \bibinfo {author} {\bibfnamefont {D.}~\bibnamefont
  {Mayou}}, \bibinfo {author} {\bibfnamefont {T.}~\bibnamefont {Li}}, \bibinfo
  {author} {\bibfnamefont {J.}~\bibnamefont {Hass}}, \bibinfo {author}
  {\bibfnamefont {A.~N.}\ \bibnamefont {Marchenkov}}, \bibinfo {author}
  {\bibfnamefont {E.~H.}\ \bibnamefont {Conrad}}, \bibinfo {author}
  {\bibfnamefont {P.~N.}\ \bibnamefont {First}}, \ and\ \bibinfo {author}
  {\bibfnamefont {W.~A.}\ \bibnamefont {de~Heer}},\ }\href {\doibase
  10.1126/science.1125925} {\bibfield  {journal} {\bibinfo  {journal}
  {Science}\ }\textbf {\bibinfo {volume} {312}},\ \bibinfo {pages} {1191}
  (\bibinfo {year} {2006})}\BibitemShut {NoStop}%
\bibitem [{\citenamefont {Blake}\ \emph {et~al.}(2008)\citenamefont {Blake},
  \citenamefont {Brimicombe}, \citenamefont {Nair}, \citenamefont {Booth},
  \citenamefont {Jiang}, \citenamefont {Schedin}, \citenamefont {Ponomarenko},
  \citenamefont {Morozov}, \citenamefont {Gleeson}, \citenamefont {Hill},
  \citenamefont {Geim},\ and\ \citenamefont {Novoselov}}]{14}%
  \BibitemOpen
  \bibfield  {author} {\bibinfo {author} {\bibfnamefont {P.}~\bibnamefont
  {Blake}}, \bibinfo {author} {\bibfnamefont {P.~D.}\ \bibnamefont
  {Brimicombe}}, \bibinfo {author} {\bibfnamefont {R.~R.}\ \bibnamefont
  {Nair}}, \bibinfo {author} {\bibfnamefont {T.~J.}\ \bibnamefont {Booth}},
  \bibinfo {author} {\bibfnamefont {D.}~\bibnamefont {Jiang}}, \bibinfo
  {author} {\bibfnamefont {F.}~\bibnamefont {Schedin}}, \bibinfo {author}
  {\bibfnamefont {L.~A.}\ \bibnamefont {Ponomarenko}}, \bibinfo {author}
  {\bibfnamefont {S.~V.}\ \bibnamefont {Morozov}}, \bibinfo {author}
  {\bibfnamefont {H.~F.}\ \bibnamefont {Gleeson}}, \bibinfo {author}
  {\bibfnamefont {E.~W.}\ \bibnamefont {Hill}}, \bibinfo {author}
  {\bibfnamefont {A.~K.}\ \bibnamefont {Geim}}, \ and\ \bibinfo {author}
  {\bibfnamefont {K.~S.}\ \bibnamefont {Novoselov}},\ }\href {\doibase
  10.1021/nl080649i} {\bibfield  {journal} {\bibinfo  {journal} {Nano Lett.}\
  }\textbf {\bibinfo {volume} {8}},\ \bibinfo {pages} {1704} (\bibinfo {year}
  {2008})}\BibitemShut {NoStop}%
\bibitem [{\citenamefont {Hernandez}\ \emph {et~al.}(2008)\citenamefont
  {Hernandez}, \citenamefont {Nicolosi}, \citenamefont {Lotya}, \citenamefont
  {Blighe}, \citenamefont {Sun}, \citenamefont {De}, \citenamefont {T.},
  \citenamefont {Holland}, \citenamefont {Byrne}, \citenamefont {Gun'Ko},
  \citenamefont {Boland}, \citenamefont {Niraj}, \citenamefont {Duesberg},
  \citenamefont {Krishnamurthy}, \citenamefont {Goodhue}, \citenamefont
  {Hutchison}, \citenamefont {Scardaci}, \citenamefont {Ferrari},\ and\
  \citenamefont {Coleman}}]{15}%
  \BibitemOpen
  \bibfield  {author} {\bibinfo {author} {\bibfnamefont {Y.}~\bibnamefont
  {Hernandez}}, \bibinfo {author} {\bibfnamefont {V.}~\bibnamefont {Nicolosi}},
  \bibinfo {author} {\bibfnamefont {M.}~\bibnamefont {Lotya}}, \bibinfo
  {author} {\bibfnamefont {F.~M.}\ \bibnamefont {Blighe}}, \bibinfo {author}
  {\bibfnamefont {Z.}~\bibnamefont {Sun}}, \bibinfo {author} {\bibfnamefont
  {S.}~\bibnamefont {De}}, \bibinfo {author} {\bibfnamefont {M.}~\bibnamefont
  {T.}}, \bibinfo {author} {\bibfnamefont {B.}~\bibnamefont {Holland}},
  \bibinfo {author} {\bibfnamefont {M.}~\bibnamefont {Byrne}}, \bibinfo
  {author} {\bibfnamefont {Y.~K.}\ \bibnamefont {Gun'Ko}}, \bibinfo {author}
  {\bibfnamefont {J.~J.}\ \bibnamefont {Boland}}, \bibinfo {author}
  {\bibfnamefont {P.}~\bibnamefont {Niraj}}, \bibinfo {author} {\bibfnamefont
  {G.}~\bibnamefont {Duesberg}}, \bibinfo {author} {\bibfnamefont
  {S.}~\bibnamefont {Krishnamurthy}}, \bibinfo {author} {\bibfnamefont
  {R.}~\bibnamefont {Goodhue}}, \bibinfo {author} {\bibfnamefont
  {J.}~\bibnamefont {Hutchison}}, \bibinfo {author} {\bibfnamefont
  {V.}~\bibnamefont {Scardaci}}, \bibinfo {author} {\bibfnamefont {A.~C.}\
  \bibnamefont {Ferrari}}, \ and\ \bibinfo {author} {\bibfnamefont {J.~N.}\
  \bibnamefont {Coleman}},\ }\href {\doibase 10.1038/nnano.2008.215} {\bibfield
   {journal} {\bibinfo  {journal} {Nat. Nanotechnol.}\ }\textbf {\bibinfo
  {volume} {3}},\ \bibinfo {pages} {563} (\bibinfo {year} {2008})}\BibitemShut
  {NoStop}%
\bibitem [{\citenamefont {Choucair}\ \emph {et~al.}(2009)\citenamefont
  {Choucair}, \citenamefont {Thordarson},\ and\ \citenamefont {Stride}}]{16}%
  \BibitemOpen
  \bibfield  {author} {\bibinfo {author} {\bibfnamefont {M.}~\bibnamefont
  {Choucair}}, \bibinfo {author} {\bibfnamefont {P.}~\bibnamefont
  {Thordarson}}, \ and\ \bibinfo {author} {\bibfnamefont {J.~A.}\ \bibnamefont
  {Stride}},\ }\href {\doibase 10.1038/nnano.2008.365} {\bibfield  {journal}
  {\bibinfo  {journal} {Nat. Nanotechnol.}\ }\textbf {\bibinfo {volume} {4}},\
  \bibinfo {pages} {30} (\bibinfo {year} {2009})}\BibitemShut {NoStop}%
\bibitem [{\citenamefont {Li}\ \emph {et~al.}(2009)\citenamefont {Li},
  \citenamefont {Cai}, \citenamefont {An}, \citenamefont {Kim}, \citenamefont
  {Nah}, \citenamefont {Yang}, \citenamefont {Piner}, \citenamefont
  {Velamakanni}, \citenamefont {Jung}, \citenamefont {Tutuc}, \citenamefont
  {Banerjee}, \citenamefont {Colombo},\ and\ \citenamefont
  {Ruoff}}]{LiScience}%
  \BibitemOpen
  \bibfield  {author} {\bibinfo {author} {\bibfnamefont {X.}~\bibnamefont
  {Li}}, \bibinfo {author} {\bibfnamefont {W.}~\bibnamefont {Cai}}, \bibinfo
  {author} {\bibfnamefont {J.}~\bibnamefont {An}}, \bibinfo {author}
  {\bibfnamefont {S.}~\bibnamefont {Kim}}, \bibinfo {author} {\bibfnamefont
  {J.}~\bibnamefont {Nah}}, \bibinfo {author} {\bibfnamefont {D.}~\bibnamefont
  {Yang}}, \bibinfo {author} {\bibfnamefont {R.}~\bibnamefont {Piner}},
  \bibinfo {author} {\bibfnamefont {A.}~\bibnamefont {Velamakanni}}, \bibinfo
  {author} {\bibfnamefont {I.}~\bibnamefont {Jung}}, \bibinfo {author}
  {\bibfnamefont {E.}~\bibnamefont {Tutuc}}, \bibinfo {author} {\bibfnamefont
  {S.~K.}\ \bibnamefont {Banerjee}}, \bibinfo {author} {\bibfnamefont
  {L.}~\bibnamefont {Colombo}}, \ and\ \bibinfo {author} {\bibfnamefont
  {R.~S.}\ \bibnamefont {Ruoff}},\ }\href {\doibase 10.1126/science.1171245}
  {\bibfield  {journal} {\bibinfo  {journal} {Science}\ }\textbf {\bibinfo
  {volume} {324}},\ \bibinfo {pages} {1312} (\bibinfo {year}
  {2009})}\BibitemShut {NoStop}%
\bibitem [{\citenamefont {Bae}\ \emph {et~al.}(2010)\citenamefont {Bae},
  \citenamefont {Kim}, \citenamefont {Lee}, \citenamefont {Xu}, \citenamefont
  {Park}, \citenamefont {Zheng}, \citenamefont {Balakrishnan}, \citenamefont
  {Lei}, \citenamefont {Ri~Kim}, \citenamefont {Song}, \citenamefont {Kim},
  \citenamefont {Kim}, \citenamefont {Ozyilmaz}, \citenamefont {Ahn},
  \citenamefont {Hong},\ and\ \citenamefont {Iijima}}]{Bae}%
  \BibitemOpen
  \bibfield  {author} {\bibinfo {author} {\bibfnamefont {S.}~\bibnamefont
  {Bae}}, \bibinfo {author} {\bibfnamefont {H.}~\bibnamefont {Kim}}, \bibinfo
  {author} {\bibfnamefont {Y.}~\bibnamefont {Lee}}, \bibinfo {author}
  {\bibfnamefont {X.}~\bibnamefont {Xu}}, \bibinfo {author} {\bibfnamefont
  {J.-S.}\ \bibnamefont {Park}}, \bibinfo {author} {\bibfnamefont
  {Y.}~\bibnamefont {Zheng}}, \bibinfo {author} {\bibfnamefont
  {J.}~\bibnamefont {Balakrishnan}}, \bibinfo {author} {\bibfnamefont
  {T.}~\bibnamefont {Lei}}, \bibinfo {author} {\bibfnamefont {H.}~\bibnamefont
  {Ri~Kim}}, \bibinfo {author} {\bibfnamefont {Y.~I.}\ \bibnamefont {Song}},
  \bibinfo {author} {\bibfnamefont {Y.-J.}\ \bibnamefont {Kim}}, \bibinfo
  {author} {\bibfnamefont {K.~S.}\ \bibnamefont {Kim}}, \bibinfo {author}
  {\bibfnamefont {B.}~\bibnamefont {Ozyilmaz}}, \bibinfo {author}
  {\bibfnamefont {J.-H.}\ \bibnamefont {Ahn}}, \bibinfo {author} {\bibfnamefont
  {B.~H.}\ \bibnamefont {Hong}}, \ and\ \bibinfo {author} {\bibfnamefont
  {S.}~\bibnamefont {Iijima}},\ }\href {\doibase 10.1038/nnano.2010.132}
  {\bibfield  {journal} {\bibinfo  {journal} {Nat. Nanotechnol.}\ }\textbf
  {\bibinfo {volume} {5}},\ \bibinfo {pages} {574} (\bibinfo {year}
  {2010})}\BibitemShut {NoStop}%
\bibitem [{\citenamefont {Miller}\ \emph {et~al.}(2009)\citenamefont {Miller},
  \citenamefont {Kubista}, \citenamefont {Rutter}, \citenamefont {Ruan},
  \citenamefont {de~Heer}, \citenamefont {First},\ and\ \citenamefont
  {Stroscio}}]{11Y}%
  \BibitemOpen
  \bibfield  {author} {\bibinfo {author} {\bibfnamefont {D.~L.}\ \bibnamefont
  {Miller}}, \bibinfo {author} {\bibfnamefont {K.~D.}\ \bibnamefont {Kubista}},
  \bibinfo {author} {\bibfnamefont {G.~M.}\ \bibnamefont {Rutter}}, \bibinfo
  {author} {\bibfnamefont {M.}~\bibnamefont {Ruan}}, \bibinfo {author}
  {\bibfnamefont {W.~A.}\ \bibnamefont {de~Heer}}, \bibinfo {author}
  {\bibfnamefont {P.~N.}\ \bibnamefont {First}}, \ and\ \bibinfo {author}
  {\bibfnamefont {J.~A.}\ \bibnamefont {Stroscio}},\ }\href {\doibase
  10.1126/science.1171810} {\bibfield  {journal} {\bibinfo  {journal}
  {Science}\ }\textbf {\bibinfo {volume} {324}},\ \bibinfo {pages} {924}
  (\bibinfo {year} {2009})}\BibitemShut {NoStop}%
\bibitem [{\citenamefont {Coraux}\ \emph {et~al.}(2008)\citenamefont {Coraux},
  \citenamefont {N`Diaye}, \citenamefont {Busse},\ and\ \citenamefont
  {Michely}}]{12Y}%
  \BibitemOpen
  \bibfield  {author} {\bibinfo {author} {\bibfnamefont {J.}~\bibnamefont
  {Coraux}}, \bibinfo {author} {\bibfnamefont {A.~T.}\ \bibnamefont {N`Diaye}},
  \bibinfo {author} {\bibfnamefont {C.}~\bibnamefont {Busse}}, \ and\ \bibinfo
  {author} {\bibfnamefont {T.}~\bibnamefont {Michely}},\ }\href {\doibase
  10.1021/nl0728874} {\bibfield  {journal} {\bibinfo  {journal} {Nano Lett.}\
  }\textbf {\bibinfo {volume} {8}},\ \bibinfo {pages} {565} (\bibinfo {year}
  {2008})}\BibitemShut {NoStop}%
\bibitem [{\citenamefont {Loginova}\ \emph {et~al.}(2009)\citenamefont
  {Loginova}, \citenamefont {Nie}, \citenamefont {Th\"urmer}, \citenamefont
  {Bartelt},\ and\ \citenamefont {McCarty}}]{13Y}%
  \BibitemOpen
  \bibfield  {author} {\bibinfo {author} {\bibfnamefont {E.}~\bibnamefont
  {Loginova}}, \bibinfo {author} {\bibfnamefont {S.}~\bibnamefont {Nie}},
  \bibinfo {author} {\bibfnamefont {K.}~\bibnamefont {Th\"urmer}}, \bibinfo
  {author} {\bibfnamefont {N.~C.}\ \bibnamefont {Bartelt}}, \ and\ \bibinfo
  {author} {\bibfnamefont {K.~F.}\ \bibnamefont {McCarty}},\ }\href {\doibase
  10.1103/PhysRevB.80.085430} {\bibfield  {journal} {\bibinfo  {journal} {Phys.
  Rev. B}\ }\textbf {\bibinfo {volume} {80}},\ \bibinfo {pages} {085430}
  (\bibinfo {year} {2009})}\BibitemShut {NoStop}%
\bibitem [{\citenamefont {Park}\ \emph {et~al.}(2010)\citenamefont {Park},
  \citenamefont {Meyer}, \citenamefont {Roth},\ and\ \citenamefont
  {Skakalova}}]{14Y}%
  \BibitemOpen
  \bibfield  {author} {\bibinfo {author} {\bibfnamefont {H.~J.}\ \bibnamefont
  {Park}}, \bibinfo {author} {\bibfnamefont {J.}~\bibnamefont {Meyer}},
  \bibinfo {author} {\bibfnamefont {S.}~\bibnamefont {Roth}}, \ and\ \bibinfo
  {author} {\bibfnamefont {V.}~\bibnamefont {Skakalova}},\ }\href {\doibase
  10.1016/j.carbon.2009.11.030} {\bibfield  {journal} {\bibinfo  {journal}
  {Carbon}\ }\textbf {\bibinfo {volume} {48}},\ \bibinfo {pages} {1088}
  (\bibinfo {year} {2010})}\BibitemShut {NoStop}%
\bibitem [{\citenamefont {Huang}\ \emph {et~al.}(2011)\citenamefont {Huang},
  \citenamefont {Ruiz-Vargas}, \citenamefont {van~der Zande}, \citenamefont
  {Whitney}, \citenamefont {Levendorf}, \citenamefont {Kevek}, \citenamefont
  {Garg}, \citenamefont {Alden}, \citenamefont {Hustedt}, \citenamefont {Zhu},
  \citenamefont {Park}, \citenamefont {McEuen},\ and\ \citenamefont
  {Muller}}]{mceuen}%
  \BibitemOpen
  \bibfield  {author} {\bibinfo {author} {\bibfnamefont {P.~Y.}\ \bibnamefont
  {Huang}}, \bibinfo {author} {\bibfnamefont {C.~S.}\ \bibnamefont
  {Ruiz-Vargas}}, \bibinfo {author} {\bibfnamefont {A.~M.}\ \bibnamefont
  {van~der Zande}}, \bibinfo {author} {\bibfnamefont {W.~S.}\ \bibnamefont
  {Whitney}}, \bibinfo {author} {\bibfnamefont {M.~P.}\ \bibnamefont
  {Levendorf}}, \bibinfo {author} {\bibfnamefont {J.~W.}\ \bibnamefont
  {Kevek}}, \bibinfo {author} {\bibfnamefont {S.}~\bibnamefont {Garg}},
  \bibinfo {author} {\bibfnamefont {J.~S.}\ \bibnamefont {Alden}}, \bibinfo
  {author} {\bibfnamefont {C.~J.}\ \bibnamefont {Hustedt}}, \bibinfo {author}
  {\bibfnamefont {Y.}~\bibnamefont {Zhu}}, \bibinfo {author} {\bibfnamefont
  {J.}~\bibnamefont {Park}}, \bibinfo {author} {\bibfnamefont {P.~L.}\
  \bibnamefont {McEuen}}, \ and\ \bibinfo {author} {\bibfnamefont {D.~A.}\
  \bibnamefont {Muller}},\ }\href {\doibase 10.1038/nature09718} {\bibfield
  {journal} {\bibinfo  {journal} {Nature}\ }\textbf {\bibinfo {volume} {469}},\
  \bibinfo {pages} {389} (\bibinfo {year} {2011})}\BibitemShut {NoStop}%
\bibitem [{\citenamefont {Lahiri}\ \emph {et~al.}(2010)\citenamefont {Lahiri},
  \citenamefont {Lin}, \citenamefont {Bozkurt}, \citenamefont {Oleynik},\ and\
  \citenamefont {Batzill}}]{Lahiri}%
  \BibitemOpen
  \bibfield  {author} {\bibinfo {author} {\bibfnamefont {J.}~\bibnamefont
  {Lahiri}}, \bibinfo {author} {\bibfnamefont {Y.}~\bibnamefont {Lin}},
  \bibinfo {author} {\bibfnamefont {P.}~\bibnamefont {Bozkurt}}, \bibinfo
  {author} {\bibfnamefont {I.~I.}\ \bibnamefont {Oleynik}}, \ and\ \bibinfo
  {author} {\bibfnamefont {M.}~\bibnamefont {Batzill}},\ }\href {\doibase
  10.1038/NNANO.2010.53} {\bibfield  {journal} {\bibinfo  {journal} {Nat.
  Nanotechnol.}\ }\textbf {\bibinfo {volume} {5}},\ \bibinfo {pages} {326}
  (\bibinfo {year} {2010})}\BibitemShut {NoStop}%
\bibitem [{\citenamefont {Yazyev}\ and\ \citenamefont
  {Louie}(2010{\natexlab{a}})}]{Yazyev}%
  \BibitemOpen
  \bibfield  {author} {\bibinfo {author} {\bibfnamefont {O.~V.}\ \bibnamefont
  {Yazyev}}\ and\ \bibinfo {author} {\bibfnamefont {S.~G.}\ \bibnamefont
  {Louie}},\ }\href {\doibase 10.1103/PhysRevB.81.195420} {\bibfield  {journal}
  {\bibinfo  {journal} {Phys. Rev. B}\ }\textbf {\bibinfo {volume} {81}},\
  \bibinfo {pages} {195420} (\bibinfo {year} {2010}{\natexlab{a}})}\BibitemShut
  {NoStop}%
\bibitem [{\citenamefont {Grantab}\ \emph {et~al.}(2010)\citenamefont
  {Grantab}, \citenamefont {Shenoy},\ and\ \citenamefont {Ruoff}}]{Grantab}%
  \BibitemOpen
  \bibfield  {author} {\bibinfo {author} {\bibfnamefont {R.}~\bibnamefont
  {Grantab}}, \bibinfo {author} {\bibfnamefont {V.~B.}\ \bibnamefont {Shenoy}},
  \ and\ \bibinfo {author} {\bibfnamefont {R.~S.}\ \bibnamefont {Ruoff}},\
  }\href {\doibase 10.1126/science.1196893} {\bibfield  {journal} {\bibinfo
  {journal} {Science}\ }\textbf {\bibinfo {volume} {330}},\ \bibinfo {pages}
  {946} (\bibinfo {year} {2010})}\BibitemShut {NoStop}%
\bibitem [{\citenamefont {Malola}\ \emph {et~al.}(2010)\citenamefont {Malola},
  \citenamefont {H\"akkinen},\ and\ \citenamefont {Koskinen}}]{Malola}%
  \BibitemOpen
  \bibfield  {author} {\bibinfo {author} {\bibfnamefont {S.}~\bibnamefont
  {Malola}}, \bibinfo {author} {\bibfnamefont {H.}~\bibnamefont {H\"akkinen}},
  \ and\ \bibinfo {author} {\bibfnamefont {P.}~\bibnamefont {Koskinen}},\
  }\href {\doibase 10.1103/PhysRevB.81.165447} {\bibfield  {journal} {\bibinfo
  {journal} {Phys. Rev. B}\ }\textbf {\bibinfo {volume} {81}},\ \bibinfo
  {pages} {165447} (\bibinfo {year} {2010})}\BibitemShut {NoStop}%
\bibitem [{\citenamefont {Carlsson}\ \emph {et~al.}(2011)\citenamefont
  {Carlsson}, \citenamefont {Ghiringhelli},\ and\ \citenamefont
  {Fasolino}}]{Carlsson}%
  \BibitemOpen
  \bibfield  {author} {\bibinfo {author} {\bibfnamefont {J.~M.}\ \bibnamefont
  {Carlsson}}, \bibinfo {author} {\bibfnamefont {L.~M.}\ \bibnamefont
  {Ghiringhelli}}, \ and\ \bibinfo {author} {\bibfnamefont {A.}~\bibnamefont
  {Fasolino}},\ }\href {\doibase 10.1103/PhysRevB.84.165423} {\bibfield
  {journal} {\bibinfo  {journal} {Phys. Rev. B}\ }\textbf {\bibinfo {volume}
  {84}},\ \bibinfo {pages} {165423} (\bibinfo {year} {2011})}\BibitemShut
  {NoStop}%
\bibitem [{\citenamefont {Yazyev}\ and\ \citenamefont
  {Louie}(2010{\natexlab{b}})}]{Yazyev2}%
  \BibitemOpen
  \bibfield  {author} {\bibinfo {author} {\bibfnamefont {O.~V.}\ \bibnamefont
  {Yazyev}}\ and\ \bibinfo {author} {\bibfnamefont {S.~G.}\ \bibnamefont
  {Louie}},\ }\href {\doibase 10.1038/nmat2830} {\bibfield  {journal} {\bibinfo
   {journal} {Nat. Mater.}\ }\textbf {\bibinfo {volume} {9}},\ \bibinfo {pages}
  {806} (\bibinfo {year} {2010}{\natexlab{b}})}\BibitemShut {NoStop}%
\bibitem [{\citenamefont {Kronberg}\ and\ \citenamefont {Wilson}(1949)}]{CSL}%
  \BibitemOpen
  \bibfield  {author} {\bibinfo {author} {\bibfnamefont {M.~L.}\ \bibnamefont
  {Kronberg}}\ and\ \bibinfo {author} {\bibfnamefont {F.~H.}\ \bibnamefont
  {Wilson}},\ }\href@noop {} {\bibfield  {journal} {\bibinfo  {journal} {J.
  Metals}\ }\textbf {\bibinfo {volume} {185}},\ \bibinfo {pages} {501}
  (\bibinfo {year} {1949})}\BibitemShut {NoStop}%
\bibitem [{\citenamefont {Valiev}\ \emph {et~al.}(2000)\citenamefont {Valiev},
  \citenamefont {Islamgaliev},\ and\ \citenamefont {Alexandrov}}]{Valiev}%
  \BibitemOpen
  \bibfield  {author} {\bibinfo {author} {\bibfnamefont {R.~Z.}\ \bibnamefont
  {Valiev}}, \bibinfo {author} {\bibfnamefont {R.~K.}\ \bibnamefont
  {Islamgaliev}}, \ and\ \bibinfo {author} {\bibfnamefont {I.~V.}\ \bibnamefont
  {Alexandrov}},\ }\href {\doibase 10.1016/S0079-6425(99)00007-9} {\bibfield
  {journal} {\bibinfo  {journal} {Prog. Mater. Sci.}\ }\textbf {\bibinfo
  {volume} {45}},\ \bibinfo {pages} {103} (\bibinfo {year} {2000})}\BibitemShut
  {NoStop}%
\bibitem [{\citenamefont {Mesaros}\ \emph {et~al.}(2010)\citenamefont
  {Mesaros}, \citenamefont {Papanikolaou}, \citenamefont {Flipse},
  \citenamefont {Sadri},\ and\ \citenamefont {Zaanen}}]{Mesaros}%
  \BibitemOpen
  \bibfield  {author} {\bibinfo {author} {\bibfnamefont {A.}~\bibnamefont
  {Mesaros}}, \bibinfo {author} {\bibfnamefont {S.}~\bibnamefont
  {Papanikolaou}}, \bibinfo {author} {\bibfnamefont {C.~F.~J.}\ \bibnamefont
  {Flipse}}, \bibinfo {author} {\bibfnamefont {D.}~\bibnamefont {Sadri}}, \
  and\ \bibinfo {author} {\bibfnamefont {J.}~\bibnamefont {Zaanen}},\ }\href
  {\doibase 10.1103/PhysRevB.82.205119} {\bibfield  {journal} {\bibinfo
  {journal} {Phys. Rev. B}\ }\textbf {\bibinfo {volume} {82}},\ \bibinfo
  {pages} {205119} (\bibinfo {year} {2010})}\BibitemShut {NoStop}%
\bibitem [{\citenamefont {Kotakoski}\ \emph {et~al.}(2011)\citenamefont
  {Kotakoski}, \citenamefont {Krasheninnikov}, \citenamefont {Kaiser},\ and\
  \citenamefont {Meyer}}]{Meyer}%
  \BibitemOpen
  \bibfield  {author} {\bibinfo {author} {\bibfnamefont {J.}~\bibnamefont
  {Kotakoski}}, \bibinfo {author} {\bibfnamefont {A.~V.}\ \bibnamefont
  {Krasheninnikov}}, \bibinfo {author} {\bibfnamefont {U.}~\bibnamefont
  {Kaiser}}, \ and\ \bibinfo {author} {\bibfnamefont {J.~C.}\ \bibnamefont
  {Meyer}},\ }\href {\doibase 10.1103/PhysRevLett.106.105505} {\bibfield
  {journal} {\bibinfo  {journal} {Phys. Rev. Lett.}\ }\textbf {\bibinfo
  {volume} {106}},\ \bibinfo {pages} {105505} (\bibinfo {year}
  {2011})}\BibitemShut {NoStop}%
\bibitem [{\citenamefont {Yazyev}(2010)}]{YazyevREP}%
  \BibitemOpen
  \bibfield  {author} {\bibinfo {author} {\bibfnamefont {O.~V.}\ \bibnamefont
  {Yazyev}},\ }\href {\doibase 10.1088/0034-4885/73/5/056501} {\bibfield
  {journal} {\bibinfo  {journal} {Rep. Prog. Phys.}\ }\textbf {\bibinfo
  {volume} {73}},\ \bibinfo {pages} {056501} (\bibinfo {year}
  {2010})}\BibitemShut {NoStop}%
\bibitem [{\citenamefont {Cervenka}\ \emph {et~al.}(2009)\citenamefont
  {Cervenka}, \citenamefont {Katsnelson},\ and\ \citenamefont
  {Flipse}}]{Cervenka}%
  \BibitemOpen
  \bibfield  {author} {\bibinfo {author} {\bibfnamefont {J.}~\bibnamefont
  {Cervenka}}, \bibinfo {author} {\bibfnamefont {M.~I.}\ \bibnamefont
  {Katsnelson}}, \ and\ \bibinfo {author} {\bibfnamefont {C.~F.~J.}\
  \bibnamefont {Flipse}},\ }\href {\doibase 10.1038/NPHYS1399} {\bibfield
  {journal} {\bibinfo  {journal} {Nat. Phys.}\ }\textbf {\bibinfo {volume}
  {5}},\ \bibinfo {pages} {840} (\bibinfo {year} {2009})}\BibitemShut {NoStop}%
\bibitem [{\citenamefont {Mart\'inez-Mart\'in}\ \emph
  {et~al.}(2010)\citenamefont {Mart\'inez-Mart\'in}, \citenamefont {Jaafar},
  \citenamefont {P\'erez}, \citenamefont {G\'omez-Herrero},\ and\ \citenamefont
  {Asenjo}}]{PRLspanish}%
  \BibitemOpen
  \bibfield  {author} {\bibinfo {author} {\bibfnamefont {D.}~\bibnamefont
  {Mart\'inez-Mart\'in}}, \bibinfo {author} {\bibfnamefont {M.}~\bibnamefont
  {Jaafar}}, \bibinfo {author} {\bibfnamefont {R.}~\bibnamefont {P\'erez}},
  \bibinfo {author} {\bibfnamefont {J.}~\bibnamefont {G\'omez-Herrero}}, \ and\
  \bibinfo {author} {\bibfnamefont {A.}~\bibnamefont {Asenjo}},\ }\href
  {\doibase 10.1103/PhysRevLett.105.257203} {\bibfield  {journal} {\bibinfo
  {journal} {Phys. Rev. Lett.}\ }\textbf {\bibinfo {volume} {105}},\ \bibinfo
  {pages} {257203} (\bibinfo {year} {2010})}\BibitemShut {NoStop}%
\bibitem {Irina} M. Sepioni, R. R. Nair, I-Ling Tsai, A. K. Geim, I. V. Grigorieva
  Europhys. Lett. {\bf 97}, 47001 (2012).
\bibitem [{\citenamefont {Edwards}\ and\ \citenamefont
  {Katsnelson}(2006)}]{Edwards2006}%
  \BibitemOpen
  \bibfield  {author} {\bibinfo {author} {\bibfnamefont {D.~M.}\ \bibnamefont
  {Edwards}}\ and\ \bibinfo {author} {\bibfnamefont {M.~I.}\ \bibnamefont
  {Katsnelson}},\ }\href {\doibase 10.1088/0953-8984/18/31/016} {\bibfield
  {journal} {\bibinfo  {journal} {J. Phys.: Condens. Matter}\ }\textbf
  {\bibinfo {volume} {18}},\ \bibinfo {pages} {7209} (\bibinfo {year}
  {2006})}\BibitemShut {NoStop}%
\bibitem [{\citenamefont {Los}\ \emph {et~al.}(2005)\citenamefont {Los},
  \citenamefont {Ghiringhelli}, \citenamefont {Meijer},\ and\ \citenamefont
  {Fasolino}}]{LCBOPII}%
  \BibitemOpen
  \bibfield  {author} {\bibinfo {author} {\bibfnamefont {J.~H.}\ \bibnamefont
  {Los}}, \bibinfo {author} {\bibfnamefont {L.~M.}\ \bibnamefont
  {Ghiringhelli}}, \bibinfo {author} {\bibfnamefont {E.~J.}\ \bibnamefont
  {Meijer}}, \ and\ \bibinfo {author} {\bibfnamefont {A.}~\bibnamefont
  {Fasolino}},\ }\href {\doibase 10.1103/PhysRevB.72.214102} {\bibfield
  {journal} {\bibinfo  {journal} {Phys. Rev. B}\ }\textbf {\bibinfo {volume}
  {72}},\ \bibinfo {pages} {214102} (\bibinfo {year} {2005})}\BibitemShut
  {NoStop}%
\bibitem [{xyz()}]{xyz2eps}%
  \BibitemOpen
  \href@noop {} {}\Eprint
  {http://arxiv.org/abs/http://sourceforge.net/projects/xyz2eps/}
  {http://sourceforge.net/projects/xyz2eps/} \BibitemShut {NoStop}%
\bibitem [{pyt()}]{python}%
  \BibitemOpen
  \href@noop {} {\enquote {\bibinfo {title} {Python software foundation, python
  programming language, version 2.6.5},}\ }\Eprint
  {http://arxiv.org/abs/http://www.python.org/} {http://www.python.org/}
  \BibitemShut {NoStop}%
\bibitem [{\citenamefont {Momma}\ and\ \citenamefont {Izumi}(2008)}]{vesta}%
  \BibitemOpen
  \bibfield  {author} {\bibinfo {author} {\bibfnamefont {K.}~\bibnamefont
  {Momma}}\ and\ \bibinfo {author} {\bibfnamefont {F.}~\bibnamefont {Izumi}},\
  }\href {\doibase 10.1107/S0021889808012016} {\bibfield  {journal} {\bibinfo
  {journal} {J. Appl. Crystallogr.}\ }\textbf {\bibinfo {volume} {41}},\
  \bibinfo {pages} {653} (\bibinfo {year} {2008})}\BibitemShut {NoStop}%
\bibitem [{\citenamefont {Fasolino}\ \emph {et~al.}(2007)\citenamefont
  {Fasolino}, \citenamefont {Los},\ and\ \citenamefont {Katsnelson}}]{NatMat}%
  \BibitemOpen
  \bibfield  {author} {\bibinfo {author} {\bibfnamefont {A.}~\bibnamefont
  {Fasolino}}, \bibinfo {author} {\bibfnamefont {J.~H.}\ \bibnamefont {Los}}, \
  and\ \bibinfo {author} {\bibfnamefont {M.~I.}\ \bibnamefont {Katsnelson}},\
  }\href {\doibase 10.1038/nmat2011} {\bibfield  {journal} {\bibinfo  {journal}
  {Nat. Mater.}\ }\textbf {\bibinfo {volume} {6}},\ \bibinfo {pages} {858}
  (\bibinfo {year} {2007})}\BibitemShut {NoStop}%
\bibitem [{\citenamefont {Los}\ \emph {et~al.}(2009)\citenamefont {Los},
  \citenamefont {Katsnelson}, \citenamefont {Yazyev}, \citenamefont
  {Zakharchenko},\ and\ \citenamefont {Fasolino}}]{PRBripple}%
  \BibitemOpen
  \bibfield  {author} {\bibinfo {author} {\bibfnamefont {J.~H.}\ \bibnamefont
  {Los}}, \bibinfo {author} {\bibfnamefont {M.~I.}\ \bibnamefont {Katsnelson}},
  \bibinfo {author} {\bibfnamefont {O.~V.}\ \bibnamefont {Yazyev}}, \bibinfo
  {author} {\bibfnamefont {K.~V.}\ \bibnamefont {Zakharchenko}}, \ and\
  \bibinfo {author} {\bibfnamefont {A.}~\bibnamefont {Fasolino}},\ }\href
  {\doibase 10.1103/PhysRevB.80.121405} {\bibfield  {journal} {\bibinfo
  {journal} {Phys. Rev. B}\ }\textbf {\bibinfo {volume} {80}},\ \bibinfo
  {pages} {121405} (\bibinfo {year} {2009})}\BibitemShut {NoStop}%
\bibitem [{\citenamefont {Zakharchenko}\ \emph {et~al.}(2009)\citenamefont
  {Zakharchenko}, \citenamefont {Katsnelson},\ and\ \citenamefont
  {Fasolino}}]{PRL2009}%
  \BibitemOpen
  \bibfield  {author} {\bibinfo {author} {\bibfnamefont {K.~V.}\ \bibnamefont
  {Zakharchenko}}, \bibinfo {author} {\bibfnamefont {M.~I.}\ \bibnamefont
  {Katsnelson}}, \ and\ \bibinfo {author} {\bibfnamefont {A.}~\bibnamefont
  {Fasolino}},\ }\href {\doibase 10.1103/PhysRevLett.102.046808} {\bibfield
  {journal} {\bibinfo  {journal} {Phys. Rev. Lett.}\ }\textbf {\bibinfo
  {volume} {102}},\ \bibinfo {pages} {046808} (\bibinfo {year}
  {2009})}\BibitemShut {NoStop}%
\bibitem [{\citenamefont {Karssemeijer}\ and\ \citenamefont
  {Fasolino}(2011)}]{LendertJan}%
  \BibitemOpen
  \bibfield  {author} {\bibinfo {author} {\bibfnamefont {L.~J.}\ \bibnamefont
  {Karssemeijer}}\ and\ \bibinfo {author} {\bibfnamefont {A.}~\bibnamefont
  {Fasolino}},\ }\href {\doibase 10.1016/j.susc.2010.10.036} {\bibfield
  {journal} {\bibinfo  {journal} {Surf. Sci.}\ }\textbf {\bibinfo {volume}
  {605}},\ \bibinfo {pages} {1611} (\bibinfo {year} {2011})}\BibitemShut
  {NoStop}%
\bibitem [{\citenamefont {Kroes}\ \emph {et~al.}(2011)\citenamefont {Kroes},
  \citenamefont {Akhukov}, \citenamefont {Los}, \citenamefont {Pineau},\ and\
  \citenamefont {Fasolino}}]{Jaap}%
  \BibitemOpen
  \bibfield  {author} {\bibinfo {author} {\bibfnamefont {J.~M.~H.}\
  \bibnamefont {Kroes}}, \bibinfo {author} {\bibfnamefont {M.~A.}\ \bibnamefont
  {Akhukov}}, \bibinfo {author} {\bibfnamefont {J.~H.}\ \bibnamefont {Los}},
  \bibinfo {author} {\bibfnamefont {N.}~\bibnamefont {Pineau}}, \ and\ \bibinfo
  {author} {\bibfnamefont {A.}~\bibnamefont {Fasolino}},\ }\href {\doibase
  10.1103/PhysRevB.83.165411} {\bibfield  {journal} {\bibinfo  {journal} {Phys.
  Rev. B}\ }\textbf {\bibinfo {volume} {83}},\ \bibinfo {pages} {165411}
  (\bibinfo {year} {2011})}\BibitemShut {NoStop}%
\bibitem [{\citenamefont {Zakharchenko}\ \emph {et~al.}(2010)\citenamefont
  {Zakharchenko}, \citenamefont {Los}, \citenamefont {Katsnelson},\ and\
  \citenamefont {Fasolino}}]{bil}%
  \BibitemOpen
  \bibfield  {author} {\bibinfo {author} {\bibfnamefont {K.~V.}\ \bibnamefont
  {Zakharchenko}}, \bibinfo {author} {\bibfnamefont {J.~H.}\ \bibnamefont
  {Los}}, \bibinfo {author} {\bibfnamefont {M.~I.}\ \bibnamefont {Katsnelson}},
  \ and\ \bibinfo {author} {\bibfnamefont {A.}~\bibnamefont {Fasolino}},\
  }\href {\doibase 10.1103/PhysRevB.81.235439} {\bibfield  {journal} {\bibinfo
  {journal} {Phys. Rev. B}\ }\textbf {\bibinfo {volume} {81}},\ \bibinfo
  {pages} {235439} (\bibinfo {year} {2010})}\BibitemShut {NoStop}%
\bibitem [{\citenamefont {Hohenberg}\ and\ \citenamefont {Kohn}(1964)}]{DFT-1}%
  \BibitemOpen
  \bibfield  {author} {\bibinfo {author} {\bibfnamefont {P.}~\bibnamefont
  {Hohenberg}}\ and\ \bibinfo {author} {\bibfnamefont {W.}~\bibnamefont
  {Kohn}},\ }\href {\doibase 10.1103/PhysRev.136.B864} {\bibfield  {journal}
  {\bibinfo  {journal} {Phys. Rev.}\ }\textbf {\bibinfo {volume} {136}},\
  \bibinfo {pages} {B864} (\bibinfo {year} {1964})}\BibitemShut {NoStop}%
\bibitem [{\citenamefont {Kohn}\ and\ \citenamefont {Sham}(1965)}]{DFT-2}%
  \BibitemOpen
  \bibfield  {author} {\bibinfo {author} {\bibfnamefont {W.}~\bibnamefont
  {Kohn}}\ and\ \bibinfo {author} {\bibfnamefont {L.~J.}\ \bibnamefont
  {Sham}},\ }\href {\doibase 10.1103/PhysRev.140.A1133} {\bibfield  {journal}
  {\bibinfo  {journal} {Phys. Rev.}\ }\textbf {\bibinfo {volume} {140}},\
  \bibinfo {pages} {A1133} (\bibinfo {year} {1965})}\BibitemShut {NoStop}%
\bibitem [{\citenamefont {Soler}\ \emph {et~al.}(2002)\citenamefont {Soler},
  \citenamefont {Artacho}, \citenamefont {Gale}, \citenamefont {Garcia},
  \citenamefont {Junquera}, \citenamefont {Ordejon},\ and\ \citenamefont
  {Sanchez-Portal}}]{SIESTA-1}%
  \BibitemOpen
  \bibfield  {author} {\bibinfo {author} {\bibfnamefont {J.~M.}\ \bibnamefont
  {Soler}}, \bibinfo {author} {\bibfnamefont {E.}~\bibnamefont {Artacho}},
  \bibinfo {author} {\bibfnamefont {J.~D.}\ \bibnamefont {Gale}}, \bibinfo
  {author} {\bibfnamefont {A.}~\bibnamefont {Garcia}}, \bibinfo {author}
  {\bibfnamefont {J.}~\bibnamefont {Junquera}}, \bibinfo {author}
  {\bibfnamefont {P.}~\bibnamefont {Ordejon}}, \ and\ \bibinfo {author}
  {\bibfnamefont {D.}~\bibnamefont {Sanchez-Portal}},\ }\href {\doibase
  10.1088/0953-8984/14/11/302} {\bibfield  {journal} {\bibinfo  {journal} {J.
  Phys.: Condens. Matter}\ }\textbf {\bibinfo {volume} {14}},\ \bibinfo {pages}
  {2745} (\bibinfo {year} {2002})}\BibitemShut {NoStop}%
\bibitem [{\citenamefont {Sanchez-Portal}\ \emph {et~al.}(2004)\citenamefont
  {Sanchez-Portal}, \citenamefont {Ordejon},\ and\ \citenamefont
  {Canadell}}]{SIESTA-2}%
  \BibitemOpen
  \bibfield  {author} {\bibinfo {author} {\bibfnamefont {D.}~\bibnamefont
  {Sanchez-Portal}}, \bibinfo {author} {\bibfnamefont {P.}~\bibnamefont
  {Ordejon}}, \ and\ \bibinfo {author} {\bibfnamefont {E.}~\bibnamefont
  {Canadell}},\ }\href@noop {} {\emph {\bibinfo {title} {Principles and
  Applications of Density functional Theory in Inorganic Chemistry II}}},\ 113\
  (\bibinfo  {publisher} {Berlin: Springer},\ \bibinfo {year} {2004})\ p.\
  \bibinfo {pages} {103–170}\BibitemShut {NoStop}%
\bibitem [{\citenamefont {Artacho}\ \emph {et~al.}(2008)\citenamefont
  {Artacho}, \citenamefont {Anglada}, \citenamefont {Dieguez}, \citenamefont
  {Gale}, \citenamefont {Garcia}, \citenamefont {Junquera}, \citenamefont
  {Martin}, \citenamefont {Ordejon}, \citenamefont {Pruneda}, \citenamefont
  {Sanchez-Portal},\ and\ \citenamefont {Soler}}]{SIESTA-3}%
  \BibitemOpen
  \bibfield  {author} {\bibinfo {author} {\bibfnamefont {E.}~\bibnamefont
  {Artacho}}, \bibinfo {author} {\bibfnamefont {E.}~\bibnamefont {Anglada}},
  \bibinfo {author} {\bibfnamefont {O.}~\bibnamefont {Dieguez}}, \bibinfo
  {author} {\bibfnamefont {J.~D.}\ \bibnamefont {Gale}}, \bibinfo {author}
  {\bibfnamefont {A.}~\bibnamefont {Garcia}}, \bibinfo {author} {\bibfnamefont
  {J.}~\bibnamefont {Junquera}}, \bibinfo {author} {\bibfnamefont {R.~M.}\
  \bibnamefont {Martin}}, \bibinfo {author} {\bibfnamefont {P.}~\bibnamefont
  {Ordejon}}, \bibinfo {author} {\bibfnamefont {J.~M.}\ \bibnamefont
  {Pruneda}}, \bibinfo {author} {\bibfnamefont {D.}~\bibnamefont
  {Sanchez-Portal}}, \ and\ \bibinfo {author} {\bibfnamefont {J.~M.}\
  \bibnamefont {Soler}},\ }\href {\doibase 10.1088/0953-8984/20/6/064208}
  {\bibfield  {journal} {\bibinfo  {journal} {J. Phys.: Condens. Matter}\
  }\textbf {\bibinfo {volume} {20}},\ \bibinfo {pages} {064208} (\bibinfo
  {year} {2008})}\BibitemShut {NoStop}%
\bibitem [{\citenamefont {Perdew}\ \emph {et~al.}(1996)\citenamefont {Perdew},
  \citenamefont {Burke},\ and\ \citenamefont {Ernzerhof}}]{GGA-PBE}%
  \BibitemOpen
  \bibfield  {author} {\bibinfo {author} {\bibfnamefont {J.~P.}\ \bibnamefont
  {Perdew}}, \bibinfo {author} {\bibfnamefont {K.}~\bibnamefont {Burke}}, \
  and\ \bibinfo {author} {\bibfnamefont {M.}~\bibnamefont {Ernzerhof}},\ }\href
  {\doibase 10.1103/PhysRevLett.77.3865} {\bibfield  {journal} {\bibinfo
  {journal} {Phys. Rev. Lett.}\ }\textbf {\bibinfo {volume} {77}},\ \bibinfo
  {pages} {3865} (\bibinfo {year} {1996})}\BibitemShut {NoStop}%
\bibitem [{\citenamefont {Junquera}\ \emph {et~al.}(2001)\citenamefont
  {Junquera}, \citenamefont {Paz}, \citenamefont {S\'anchez-Portal},\ and\
  \citenamefont {Artacho}}]{DZP-NAO}%
  \BibitemOpen
  \bibfield  {author} {\bibinfo {author} {\bibfnamefont {J.}~\bibnamefont
  {Junquera}}, \bibinfo {author} {\bibfnamefont {O.}~\bibnamefont {Paz}},
  \bibinfo {author} {\bibfnamefont {D.}~\bibnamefont {S\'anchez-Portal}}, \
  and\ \bibinfo {author} {\bibfnamefont {E.}~\bibnamefont {Artacho}},\ }\href
  {\doibase 10.1103/PhysRevB.64.235111} {\bibfield  {journal} {\bibinfo
  {journal} {Phys. Rev. B}\ }\textbf {\bibinfo {volume} {64}},\ \bibinfo
  {pages} {235111} (\bibinfo {year} {2001})}\BibitemShut {NoStop}%
\bibitem [{\citenamefont {Troullier}\ and\ \citenamefont
  {Martins}(1991)}]{pseudopotentials}%
  \BibitemOpen
  \bibfield  {author} {\bibinfo {author} {\bibfnamefont {N.}~\bibnamefont
  {Troullier}}\ and\ \bibinfo {author} {\bibfnamefont {J.~L.}\ \bibnamefont
  {Martins}},\ }\href {\doibase 10.1103/PhysRevB.43.1993} {\bibfield  {journal}
  {\bibinfo  {journal} {Phys. Rev. B}\ }\textbf {\bibinfo {volume} {43}},\
  \bibinfo {pages} {1993} (\bibinfo {year} {1991})}\BibitemShut {NoStop}%
\bibitem [{\citenamefont {Kleinman}\ and\ \citenamefont
  {Bylander}(1982)}]{nonlocal-form}%
  \BibitemOpen
  \bibfield  {author} {\bibinfo {author} {\bibfnamefont {L.}~\bibnamefont
  {Kleinman}}\ and\ \bibinfo {author} {\bibfnamefont {D.~M.}\ \bibnamefont
  {Bylander}},\ }\href {\doibase 10.1103/PhysRevLett.48.1425} {\bibfield
  {journal} {\bibinfo  {journal} {Phys. Rev. Lett.}\ }\textbf {\bibinfo
  {volume} {48}},\ \bibinfo {pages} {1425} (\bibinfo {year}
  {1982})}\BibitemShut {NoStop}%
\bibitem [{\citenamefont {Monkhorst}\ and\ \citenamefont
  {Pack}(1976)}]{Monkhorst-Pack}%
  \BibitemOpen
  \bibfield  {author} {\bibinfo {author} {\bibfnamefont {H.~J.}\ \bibnamefont
  {Monkhorst}}\ and\ \bibinfo {author} {\bibfnamefont {J.~D.}\ \bibnamefont
  {Pack}},\ }\href {\doibase 10.1103/PhysRevB.13.5188} {\bibfield  {journal}
  {\bibinfo  {journal} {Phys. Rev. B}\ }\textbf {\bibinfo {volume} {13}},\
  \bibinfo {pages} {5188} (\bibinfo {year} {1976})}\BibitemShut {NoStop}%
\bibitem [{\citenamefont {Saito}\ \emph {et~al.}(1988)\citenamefont {Saito},
  \citenamefont {Dresselhaus},\ and\ \citenamefont {Dresselhaus}}]{nanotubes}%
  \BibitemOpen
  \bibfield  {author} {\bibinfo {author} {\bibfnamefont {R.}~\bibnamefont
  {Saito}}, \bibinfo {author} {\bibfnamefont {G.}~\bibnamefont {Dresselhaus}},
  \ and\ \bibinfo {author} {\bibfnamefont {M.~S.}\ \bibnamefont
  {Dresselhaus}},\ }\href@noop {} {\emph {\bibinfo {title} {Physical Properties
  of Carbon Nanotubes}}}\ (\bibinfo  {publisher} {Imperial College Press},\
  \bibinfo {year} {1988})\BibitemShut {NoStop}%
\bibitem [{\citenamefont {Hirth}\ and\ \citenamefont
  {Lothe}(1982)}]{HirthLother}%
  \BibitemOpen
  \bibfield  {author} {\bibinfo {author} {\bibfnamefont {J.}~\bibnamefont
  {Hirth}}\ and\ \bibinfo {author} {\bibfnamefont {J.}~\bibnamefont {Lothe}},\
  }\href@noop {} {\emph {\bibinfo {title} {Theory of Dislocations}}}\ (\bibinfo
   {publisher} {Wiley},\ \bibinfo {year} {1982})\ p.\ \bibinfo {pages}
  {870}\BibitemShut {NoStop}%
\bibitem [{\citenamefont {Zakharchenko}\ \emph {et~al.}(2011)\citenamefont
  {Zakharchenko}, \citenamefont {Fasolino}, \citenamefont {Los},\ and\
  \citenamefont {Katsnelson}}]{JCM2011}%
  \BibitemOpen
  \bibfield  {author} {\bibinfo {author} {\bibfnamefont {K.~V.}\ \bibnamefont
  {Zakharchenko}}, \bibinfo {author} {\bibfnamefont {A.}~\bibnamefont
  {Fasolino}}, \bibinfo {author} {\bibfnamefont {J.~H.}\ \bibnamefont {Los}}, \
  and\ \bibinfo {author} {\bibfnamefont {M.~I.}\ \bibnamefont {Katsnelson}},\
  }\href {\doibase 10.1088/0953-8984/23/20/202202} {\bibfield  {journal}
  {\bibinfo  {journal} {J. Phys.: Condens. Matter}\ }\textbf {\bibinfo {volume}
  {23}},\ \bibinfo {pages} {202202} (\bibinfo {year} {2011})}\BibitemShut
  {NoStop}%
\bibitem [{\citenamefont {L\'opez-Sancho}\ \emph {et~al.}(2009)\citenamefont
  {L\'opez-Sancho}, \citenamefont {de~Juan},\ and\ \citenamefont
  {Vozmediano}}]{VozmedPRB}%
  \BibitemOpen
  \bibfield  {author} {\bibinfo {author} {\bibfnamefont {M.~P.}\ \bibnamefont
  {L\'opez-Sancho}}, \bibinfo {author} {\bibfnamefont {F.}~\bibnamefont
  {de~Juan}}, \ and\ \bibinfo {author} {\bibfnamefont {M.~A.~H.}\ \bibnamefont
  {Vozmediano}},\ }\href {\doibase 10.1103/PhysRevB.79.075413} {\bibfield
  {journal} {\bibinfo  {journal} {Phys. Rev. B}\ }\textbf {\bibinfo {volume}
  {79}},\ \bibinfo {pages} {075413} (\bibinfo {year} {2009})}\BibitemShut
  {NoStop}%
\bibitem [{\citenamefont {Yazyev}\ and\ \citenamefont
  {Katsnelson}(2008)}]{YazyevEdgeMagnetism}%
  \BibitemOpen
  \bibfield  {author} {\bibinfo {author} {\bibfnamefont {O.~V.}\ \bibnamefont
  {Yazyev}}\ and\ \bibinfo {author} {\bibfnamefont {M.~I.}\ \bibnamefont
  {Katsnelson}},\ }\href {\doibase 10.1103/PhysRevLett.100.047209} {\bibfield
  {journal} {\bibinfo  {journal} {Phys. Rev. Lett.}\ }\textbf {\bibinfo
  {volume} {100}},\ \bibinfo {pages} {047209} (\bibinfo {year}
  {2008})}\BibitemShut {NoStop}%
\bibitem [{\citenamefont {Bhandary}\ \emph {et~al.}(2010)\citenamefont
  {Bhandary}, \citenamefont {Eriksson}, \citenamefont {Sanyal},\ and\
  \citenamefont {Katsnelson}}]{EdgeHydrogen}%
  \BibitemOpen
  \bibfield  {author} {\bibinfo {author} {\bibfnamefont {S.}~\bibnamefont
  {Bhandary}}, \bibinfo {author} {\bibfnamefont {O.}~\bibnamefont {Eriksson}},
  \bibinfo {author} {\bibfnamefont {B.}~\bibnamefont {Sanyal}}, \ and\ \bibinfo
  {author} {\bibfnamefont {M.~I.}\ \bibnamefont {Katsnelson}},\ }\href
  {\doibase 10.1103/PhysRevB.82.165405} {\bibfield  {journal} {\bibinfo
  {journal} {Phys. Rev. B}\ }\textbf {\bibinfo {volume} {82}},\ \bibinfo
  {pages} {165405} (\bibinfo {year} {2010})}\BibitemShut {NoStop}%
\end{thebibliography}

%

\end{document}